\documentclass[11pt]{article}

\usepackage[margin=1in]{geometry}
\usepackage{amsmath,amssymb}
\usepackage{graphicx}
\usepackage{booktabs}
\usepackage{multirow}
\usepackage{microtype}
\usepackage{xcolor}
\usepackage[hidelinks]{hyperref}
\usepackage{caption}
\usepackage{subcaption}
\usepackage{url}

\graphicspath{{figures/}}

\newcommand{\rbe}{r_{\mathrm{be}}}
\newcommand{\bfast}{\beta_{\mathrm{fast}}}
\newcommand{\bslow}{\beta_{\mathrm{slow}}}
\newcommand{\xhalf}{x_{1/2}}

\title{TerraceMoE: A Cost Model for Hierarchical MoE\\
All-to-All Communication}

\author{%
Weicheng Xue\thanks{Correspondence: \texttt{weich97@vt.edu}} \quad
Bingqiang Wang \quad
Li Yuan \quad
Huihui Zhou \quad
Yonghong Tian%
}

\date{}

\hypersetup{
  pdftitle={TerraceMoE: A Cost Model for Hierarchical MoE All-to-All Communication},
  pdfauthor={Weicheng Xue, Bingqiang Wang, Li Yuan, Huihui Zhou, Yonghong Tian}
}

\begin{document}
\maketitle

\begin{abstract}
Hierarchical two-hop dispatch can reduce slow-fabric traffic in expert-parallel
Mixture-of-Experts training, but it adds a second collective and an arrival-side
operator chain. We present a cost model for screening that trade at the
\emph{communication-call} level, bounded by validation gates that withdraw a
capability in code when they fail rather than reporting a caveat. At a reference geometry with 16 groups of 8
ranks, $q=3$, $H=2048$, and 4096 tokens per rank, the corrected effective breakeven
hierarchy ratio is 3.98 for the measured PyTorch arrival chain, 1.49 for a
hypothetical fused target, and 1.10 at zero implementation overhead. These are
ratio-only sensitivity results, not deployment predictions: platform A measures 1.03,
platform B has no separated fast/slow measurement, and neither machine measured here
reaches the hierarchical regime. Four communication-level corpora pass their gates; a drift
probe and the step-level gate fail. The latter failure is enforced in code, so we make
no training-throughput prediction. The enabling routing constraint fixes per-token
fan-out and per-selected-group quota, while aggregate per-peer counts remain
data-dependent. Its measured validation-loss cost is small but nonzero
(+0.0034 nats); downstream equivalence is reported with incomplete estimator
provenance and is therefore not independently reconstructible from the artifact.
Code, calibration constants and the validation gates are at
\url{https://github.com/weich97/TerraceMoE-simulator}.
\end{abstract}

\section{Introduction}
\label{sec:intro}

A cluster is summarised by its ratio of intra-group to inter-group bandwidth, the ratio is compared against a byte-counting threshold, and whether a hierarchical dispatch pays is taken to follow. The ratio is read off the fabric. It is not a property of the fabric. It moves with the scale-out bandwidth per accelerator a deployment buys, so one rack has two ratios and which one it has was purchased. The threshold it is compared against moves with how the arrival chain is written. And once a single fast domain holds the whole expert-parallel group, the comparison does not answer worse, it stops applying, which is where the hardware is going. Three ways for a deployment to be misclassified, none of them recoverable from the ratio a published comparison quotes.

Sparsity is how frontier language models buy parameter count without paying for it in
arithmetic. A Mixture-of-Experts layer activates a few experts per token, so capacity
grows with the expert count while arithmetic does not. Those parameters are spread
across accelerators, so expert parallelism at that scale is the mechanism rather than
an option. The bill arrives in the inner loop. Every routed token must reach its
expert's device and come back, so an all-to-all sits in every MoE layer of every step,
and in production it is often the largest single component of step
time~\cite{li2023lina,jin2026megascalemoe}. Sparsity turns a memory problem into a
communication problem.

A second trend has been reshaping the machine underneath. Interconnects have stopped
being flat. Accelerators are packaged into high-bandwidth domains, a node and
increasingly a rack, inside which links are fast and uniform, and those domains are
joined by a fabric several times slower; dozens of accelerators now share one domain
on shipping hardware~\cite{deepseekv3hw}. A collective crossing that boundary is
priced almost entirely by the slow side, whatever the peak numbers inside the domain
say.

The two trends meet at the dispatch collective, and the meeting suggests an obvious
response. Slow-side traffic in a flat dispatch is redundant: a token whose experts
land in several groups sends one copy per expert across the boundary where one copy
per group would do. Send the deduplicated copy to a representative rank of each target
group and let it scatter inside the group over the fast links. The shape is deployed
rather than hypothetical~\cite{deepseekv3,deepseekv3hw,deepep2025}.

It is not obviously right. Two hops pay two collective fixed costs where one hop paid
one. The bytes taken off the slow side reappear on the fast side, cheaper but not
free. A deduplicated message must also be unpacked on arrival into a local
permutation, which costs real time and grows with hidden width. Whether the trade pays
is therefore a property of a particular machine and a particular implementation, not
of the idea, and the deciding terms are invisible in a topology diagram.

The field settles such questions by building both paths and measuring
them~\cite{li2023lina,jin2026megascalemoe}. That is expensive, and asymmetric in a way
that matters more than its size. A positive verdict is audited by the deployment that
follows it: the system ships and step time says whether the restructuring paid. A
negative verdict is never audited, because nobody builds what was rejected, and no
measurement contradicts a system that does not exist. The answers that would save the
most engineering are the ones nobody can check, which is why a practitioner is least
willing to take one on trust.

Computing the verdict instead is not out of reach, because the raw material is cheap.
The primitives a dispatch would be assembled from, a collective at a given message
size, a gather over a given tensor shape, can each be timed directly on hardware that
already exists; the composite, at a world size and a bandwidth ratio nobody in the
room has, cannot. Composition, not measurement, is the hard step.

The models that perform it are not built to be disbelieved. MoE communication models
price the deduplication and report the speedups they
found~\cite{xmoe2025,lin2025hiermoe}; the simulators that precede this one validate by
reporting an accuracy
figure~\cite{isaev2023calculon,duan2024proteus,wang2025simai,agrawal2024vidur}. An
accuracy figure is a summary, and a summary cannot withdraw a capability: it says how
a model did on average over a corpus its authors chose, not what it may no longer
predict when a check fails. For a verdict of yes that gap is survivable, since the
build finds out. For a verdict of no it is the whole problem.
Section~\ref{sec:related} places the model among its predecessors.

The two halves of the question come from different places. Cheap primitives set
against an expensive build ask whether the composite can be predicted; the audit
asymmetry demands more than an average behind the prediction. Can the verdict on a
dispatch restructuring be predicted from microbenchmarks of the primitives it would be
assembled from, for machines and geometries nobody has built, and can that prediction
be made accountable enough that a verdict of no is actionable on its own?

At the communication-call level the first half clears the stated gates within the
measured coverage; the second half is what makes that limited answer usable. Applied
as a ratio sensitivity study, the model shows that hierarchy ratio does not determine
the outcome on its own. Accountability comes
from gates bound in code to what the model may output, so a failure removes a
capability rather than adds a caveat. Two fail here and are left standing.

One end-to-end build of both paths supplies a validation instance. The
communication-call model scores two-hop as a loss at the measured ratio, while the
step-time bed agrees at five of the six geometries where a direction can be scored;
the sixth is reported as a measured miss, scored against the
criterion's refutation rule, itself post hoc and labeled as such, rather than averaged
away (Section~\ref{sec:e2e}). Because the step-level gate fails, this agreement is
diagnostic rather than a licensed throughput prediction.

\paragraph{Contributions.}
\begin{enumerate}
\item A cost model for MoE all-to-all whose functional form retains acceptable fit on
a second machine after machine-specific refitting, governed by gates whose failure removes a
capability rather than by an accuracy figure, with every threshold and reporting rule
assigned its own registration status rather than a blanket claim
(Sections~\ref{sec:model}
and~\ref{sec:validation}). Two gates fail; the step-level failure is an interlock in
code, so no step-level prediction appears anywhere below.

\item A ratio-only sensitivity analysis that separates measured machine constants from
synthetic hierarchy-ratio scenarios (Section~\ref{sec:applicability}). Provisioning,
arrival-chain implementation, and fast-domain size all change the interpretation of a
quoted ratio; the model prices the first two and checks the third as a precondition.
\item A measured characterization of the routing constraint that makes two-hop
expressible (Section~\ref{sec:troute}). It bounds per-token fan-out and fixes the
per-selected-group quota. The validation-loss effect is resolved and small, while the
downstream, load, and step-time evidence carries the limitations stated there.
\end{enumerate}

The simulator is TerraceMoE; the methods it prices share the
prefix, T-Route for the routing constraint and T-A2A for the dispatch it makes
expressible. Everything, gates included, is at
\url{https://github.com/weich97/TerraceMoE-simulator}. Both failing gates run against
a reader's own measurements, so the failure is falsifiable on hardware we have never
seen (Section~\ref{sec:repro}).

Only machine A has a measured hierarchy ratio, 1.03; machine B does not separate its
fast and slow levels. They supply fit and consistency evidence, not observations of
hierarchical clusters. Every
statement here about higher ratios is extrapolation from the model rather than an
observation, labeled as such. Compute is nowhere added to communication; the two
overlap by an amount these measurements do not resolve (Section~\ref{sec:limits}).

\section{Problem Formulation}
\label{sec:problem}

We use the standard expert-parallel dispatch and combine
formulation~\cite{lepikhin2021gshard,fedus2022switch}: with expert parallelism degree
$EP$ and $E$ experts spread evenly over the ranks, a rank holding $T$ tokens with
top-$k$ routing emits $Tk$ payload rows per MoE layer, each row $H$ elements wide.
One-hop dispatch is a single all-to-all over the whole $EP$ group. Every row crosses
whatever link separates its source from its destination.

The links those rows cross are not all alike. We describe a machine by three
communication levels: a fast level inside a group of $R$ ranks, a slow level between
groups, and the full fabric spanning $N_g \times R = EP$ ranks. Each level carries two
quantities rather than one. The first is bandwidth, and the \emph{hierarchy ratio}
$\bfast / \bslow$ formed from the fast and slow figures, measured at the message sizes
the workload actually uses, is what everyone reaches for first. The second is the cost
of issuing a collective at all, which does not shrink with the bytes carried and
therefore cancels out of no ratio. A machine profile that keeps only the first
quantity is the abstraction this paper puts under test, and one of the results below
is that it is not sufficient: the implementation of the arrival chain moves the
threshold the ratio must clear as far as a change of interconnect would
(Section~\ref{sec:applicability}).

The dispatch that exploits those levels runs in two hops. Hop A sends, for each token
and each target group, a single copy to the representative rank of that group, chosen
as a pure function of the sender's index so no negotiation is needed. Hop B scatters
those rows inside the group to the ranks that hold the selected experts. Per token and
target group, the slow-side payload drops from $q = k/M$ rows to one row, at the cost
of $q(1 - 1/R)$ extra rows on the fast side and one additional collective.
Equation~\ref{eq:twohop} prices the whole chain.

Two-hop dispatch is only expressible with a bound on how many groups a token can
reach. We use two constraints together, neither of them ours: each token's experts are
confined to $M$ of the $N_g$ groups, which is device-limited routing~\cite{deepseekv2}
and its node-limited successor~\cite{deepseekv3}, and exactly $k/M$ experts are
selected within each chosen group, a per-group quota adapted from grouped-expert
routing~\cite{tang2025pangumoge}, which imposes it on every group rather than only on
the $M$ selected ones. The first bounds cross-group fan-out at compile time; the
second fixes the number of expert rows a token contributes to each selected group.
Aggregate per-peer messages are sums over data-dependent routing decisions and are
not constant size. Section~\ref{sec:troute}
measures what these constraints cost.

What a deployment has to decide is not the size of any of these terms but the sign of
their sum. We write $G$ for the ratio of one-hop to two-hop cost, so that $G > 1$
means the restructuring is worth doing and $G < 1$ means it is not, and we carry the
quantity at two levels that must not be conflated. At the communication level $G$
compares dispatch call time alone, which is what a model assembled from collectives
can predict. At the step level it compares whole training steps, which is what a
deployment experiences, and which additionally involves overlap with compute that
these measurements do not resolve. The two coincide only where dispatch is neither
overlapped nor amortised. Section~\ref{sec:validation} reports that the model earns
the first level and fails the second, and the failure is what confines every
prediction in this paper to dispatch call time.

Setting $G = 1$ turns the decision into a threshold on the machine. The workload terms
collapse into a single number, the breakeven hierarchy ratio $\rbe$ derived in
Section~\ref{sec:breakeven}, and a machine is worth restructuring for when its
hierarchy ratio exceeds that threshold. This is the form in which the question is
usually asked and answered, one number about the fabric set against one number about
the workload, and its sufficiency is what is at issue here: $\rbe$ counts bytes, while
the fixed costs and the arrival chain named above enter the real comparison and move
the threshold without moving the ratio. Two machines can therefore share a hierarchy
ratio and disagree on the verdict, and one machine can change its verdict without any
change to its fabric.

Answering the question for a given deployment thus requires the terms a ratio omits,
at the world size, message size and hidden width that deployment will run at. Each
term is separately measurable on hardware that already exists; the composite is not
measurable at all until the dispatch has been built, which is the expense the decision
is trying to avoid. The rest of this paper assembles the composite from the parts,
fixes in advance what would count as failing to do so
(Section~\ref{sec:design}), and reports where the result may and may not be believed.

\section{Experimental Design}
\label{sec:design}

This section states the machines, corpora, estimands, thresholds, and registration
status used below. Some rules were prespecified before their corresponding campaign;
others were amended or added after data were seen and are labeled as such. We retain
the distinction between a prediction and a reading taken after the
fact~\cite{nosek2018prereg} rather than applying a blanket preregistration label.

\subsection{Machines}
\label{sec:machines}

Two machines supply the model's constants. Table~\ref{tab:machines} gives the measured
quantities that place each of them on the axis the applicability criterion is stated
over. Machine A is the one on which we could close the loop end to end, from
primitives through prediction to a measured step time. No machine measured here supplies
a hierarchy ratio in the regime the criterion is about, which is the boundary
condition on every extrapolated statement here.

\begin{table}[!ht]
\centering
\footnotesize
\begin{tabular}{@{}llcccc p{2.2cm}@{}}
\toprule
Machine & Topology & $R$ & $\bfast$ & $\bslow$ & $\bfast/\bslow$ & Role \\
\midrule
A & unified-fabric supernode & 8 & 122.6~GB/s & 119.5~GB/s & \textbf{1.03} &
constants, verdict bed, Q1 \\
B & same family, unified fabric & 8 & \multicolumn{2}{c}{111.9~GB/s, not separated} &
not measured & fit/consistency check \\
\bottomrule
\end{tabular}
\caption{The two machines the model was instantiated on and validated against.
Bandwidths are measured at the per-peer message
sizes the MoE workload uses, not at the largest size the fabric supports. Machine A's
cross-node and intra-node bandwidths differ by 2.6\%. Machine B's corpus never
isolates its intra-node level, so a single asymptotic bandwidth is fitted for all
three levels and its hierarchy ratio was never measured. That is a real limit on what
B can be used for: it checks whether the model \emph{form} remains usable after
$\alpha$, flat $\beta$, and $\xhalf$ are fitted from C3. The fast/slow level split
and implementation-side constants were not measured on B and remain assumptions;
B is not a second observation of a hierarchy.}
\label{tab:machines}
\end{table}

\subsection{Corpus register}
\label{sec:corpora}

Six corpora are scored against a gate. Table~\ref{tab:corpora} states for each of them
what was measured, how often, when relative to the freeze of the constants in
Table~\ref{tab:constants}, and which constants were fitted on it, which is what
determines whether a later gate on that corpus is a fit-quality check or a prediction.
These labels are used throughout; no result below is identified by an ordinal.

\begin{table}[t]
\centering
\footnotesize
\begin{tabular}{@{}l p{3.05cm} p{2.15cm} p{2.35cm} p{2.55cm} p{2.15cm}@{}}
\toprule
ID & What & Machine, worlds & Size, reps & Fitted on it & Role \\
\midrule
C1 & one-hop and two-hop dispatch, measured directly & A, five token
tiers & pooled medians & $\alpha$, $\beta_\infty$ & fit, Tier-1 \\
\addlinespace
C2 & all-to-all size sweep & A & 6 targets & none & Tier-1b \\
\addlinespace
C3 & all-to-all size sweep & B & 44 targets, 19 sizes & B's constants,
$\xhalf$ & fit, Tier-1b \\
\addlinespace
C4 & all-to-all size sweep & A, world 16 & 14 targets, 6 reps & none,
post-freeze & holdout, Tier-1b \\
\addlinespace
C5 & all-to-all size sweep & A, world 8 & 13 sizes, 12 reps & none & Tier-1b \\
\addlinespace
G1 & step time, three axes & A, 7 geometries & 1 calibration, 6 holdout & one overlap
parameter & Tier-2, verdict \\
\bottomrule
\end{tabular}
\caption{The six scored corpora. C2 to C4 are scored as a conjunction, 64 targets in
total; C5 is scored separately. C4 is at a world machine A had never been scored at
before. G1 varies micro-batch size, node count, and cross-group fan-out;
the calibration geometry is the base tier, 16 groups of 8 at $k=6$, $M=2$ and one
micro-batch, and the other six are holdouts for both Tier-2 and the end-to-end verdict
of Section~\ref{sec:e2e}. The size sweep that resolves $\xhalf$ is C3's own, so for
that one parameter C3 is a consistency check and not a prediction; C4 carries the
prediction claim instead.}
\label{tab:corpora}
\end{table}

Five further threads support the model without being scored by a gate. A1 splits
$\alpha$ into host overhead and network latency, scanning $N$ from 1 to 1024
back-to-back collectives, 256~B to 16~MB, worlds 2 through 16 on two nodes, collected
after the constants were frozen (Section~\ref{sec:launch}). A2 carries the arrival
chain off its calibration point: nine row counts and four hidden widths on two nodes,
thirty iterations each (Section~\ref{sec:chainscaling}). Q1 prices the routing
constraint on model quality rather than on time: a 13.14B-parameter model, four
routing modes, four seeds per mode, 62.9B tokens per arm on machine A
(Section~\ref{sec:troute}). R1 bounds the compute term with a bf16 square GEMM
roofline at 256 samples per size across two campaigns, and R2 fixes how that curve is
indexed by a non-square shape from eight expert-FFN shapes on each of two nodes with
same-session square references (Section~\ref{sec:compute}).

\subsection{Measurement protocol and instrument resolution}
\label{sec:protocol}

Every target in this paper is a pooled median over repeated runs rather than a single
run, following the reporting discipline for nondeterministic measurements of Hoefler
and Belli~\cite{hoefler2015benchmarking}, because the instrument drifts by more than
the effects the gates are asked to resolve. Three back-to-back runs of the two-hop benchmark at 2048 tokens per rank
returned 0.648, 0.772 and 0.711~ms, a spread of 19\% about the median, and that is the
figure we mean by run-to-run drift throughout. Warmup iterations are discarded and not
pooled. \emph{The resolution of the
instrument must exceed the effect being measured}, and that applies to validation
gates as much as to experiments; it is why the Tier-1 threshold is set where
Section~\ref{sec:gatereg} sets it, and it is the reason two independent readings of
$\alpha$ can disagree by 15\% without either being wrong.

On the end-to-end bed of G1 the estimand is the step-level $G$ of
Section~\ref{sec:problem}. Its run-to-run spread on this bed is 0.6\% to
0.8\%, measured on the two geometries that were run three times each.

\subsection{Gates, thresholds, and what a failure removes}
\label{sec:gatereg}

A gate whose failure carries no consequence is decoration. Each gate below is bound to
a machine-checked constraint on what the model may output, so that a failure removes a
capability rather than adding a caveat. Table~\ref{tab:gatereg} gives, for each gate,
its estimand, its threshold, how that threshold was arrived at, whether it was
registered before its targets existed, and what its failure would remove. Outcomes are
reported separately, in Table~\ref{tab:gateresults}.

Estimands are defined once here. Tier-1 and Tier-1b score relative error of predicted
against measured collective time, per target, reported as a median and a signed median
over the corpus. Tier-2 scores mean absolute error of the predicted step-time ratio
over the six holdouts of G1, and sign agreement over the five holdouts where a sign
is defined. The sixth, $k=8$ with $M=2$, is excluded from the sign count and from it
alone: its measured ratio is 0.9935, inside the run-to-run spread of unity, so there is
no direction to get right. It remains in the six-holdout MAE.

\begin{table}[!ht]
\centering
\footnotesize
\begin{tabular}{@{}p{1.5cm} p{3.2cm} p{3.4cm} p{4.2cm}@{}}
\toprule
Gate, corpus & Estimand and threshold & How the threshold was set & Registration
status, and what a failure removes \\
\midrule
Tier-1, C1 & median relative error $\leq 20\%$, worst $\leq 35\%$, crossover inside
the measured window
& Set from the instrument, then checked against the decision. A 20\% relative error
is large enough to move a ratio-only breakeven materially, so the released code
propagates it rather than converting it into a target-platform class verdict.
& \textbf{Prespecified before C1 collection}; no timestamped public registration is
available. Failure removes every
communication-level prediction. \\
\addlinespace
Tier-1b, C2 to C4 & per machine: median $\leq 12\%$, $\leq 1$ in 5 over $35\%$, signed
median within $\pm 8\%$
& The tighter 12\% narrows the uncertainty band. The bias bound is what
a model carrying no contention term can honestly promise against corpora that contain
contended runs.
& \textbf{Post hoc regression guard} for C2 and C3, written after those targets
existed. C4 is an \textbf{out-of-sample prediction}: collected after the freeze, with
nothing fitted to it. Failure removes the claim that the same functional form retains
acceptable fit after machine-specific refitting. \\
\addlinespace
Tier-1b, C5 & same & same & \textbf{Post hoc regression guard.} Scored separately from
the C2 to C4 conjunction. \\
\addlinespace
Tier-2, G1 & MAE $\leq 0.025$ over six holdouts, $\geq 4/6$ within $\pm 0.035$, signs
correct
& Not derived in ratio units, because a failing gate never had to be defended at its
margin. The nearest thing the paper carries is in Section~\ref{sec:compute}, where an
index-rule choice moves the effective breakeven by under 2.1\%.
& \textbf{Prespecified before G1 collection}; no timestamped public registration is
available. \emph{Failure removes
step-level prediction from the model entirely}, enforced in code. \\
\bottomrule
\end{tabular}
\caption{Gate design. Status takes one of three values: prespecified before the
corresponding campaign, post hoc regression guard, or out-of-sample prediction.
\emph{Prespecified} is used instead of \emph{preregistered} because the artifact
contains no timestamped public registration.}
\label{tab:gatereg}
\end{table}

\subsection{Falsification conditions}
\label{sec:falsification}

Each claim below is stated with the observation that would refute it.

\emph{The cost model's accuracy} is refuted by Tier-1 or Tier-2, above.

\emph{The routing constraint has acceptably small quality cost} is refuted if the
paired holdout validation loss delta against the unconstrained control exceeds the
prespecified no-loss margin of
0.1 nats. That margin is one order of magnitude above the within-arm standard
deviation of the holdout probe, so an effect at the margin would be one a practitioner
would act on. Secondary endpoints in the protocol are downstream accuracy on two
benchmarks under two one-sided tests
(TOST)~\cite{schuirmann1987tost,lakens2017equivalence} at $\pm 1.0$ percentage points, expert load entropy no lower
than the control, and step time. Every arm runs four seeds and 62.9B tokens, and all
intervals are 90\% and paired by seed.

\emph{A one-sided transport would pay} was refuted by a prespecified criterion of
$\geq 1.15\times$ against collective all-to-all on at least two valid size tiers
(Section~\ref{sec:onesided}).

\emph{The applicability criterion itself} had no scoring rule, which is a gap, because
the criterion is the paper's operative output. We state one, and we state its status:
it is \textbf{post hoc}, written after the seven geometries of G1 were measured, so it
is a rule for replicators rather than a preregistration we can claim credit for.
\emph{The criterion is refuted on a machine if any geometry whose predicted verdict is
loss measures $G$ above 1 by more than the run-to-run spread of $G$ on that bed.}
Section~\ref{sec:e2e} scores G1 against it, including the one geometry that measures
1.0355.

\subsection{Reporting policy and protocol amendments}
\label{sec:policy}

Every corpus we scored against a gate is reported in Section~\ref{sec:validation},
passing or failing, and no constant was retuned to clear a gate. A model whose
failures are not visible cannot be calibrated by anyone else.

Four protocol changes were made after data were seen, and they share a shape worth
naming: in each case the instrument, not the result, turned out to be the thing that
needed fixing. On the downstream axis of Q1 a single-checkpoint read gave an estimate
whose interval touched the equivalence boundary, so two further readpoints were added
and the analysis was demoted from a point estimate to an equivalence claim
(Section~\ref{sec:troute}). The corpus behind the $\xhalf$ lower bound was repeated
six times and medians taken, which dissolved a two-sided admissibility interval into a
one-sided bound (Section~\ref{sec:uncertainty}). And the overlap families of
Section~\ref{sec:tier2} were enumerated after Tier-2 had already failed, so they are
exploratory by construction: no candidate set and no stopping rule were prespecified,
the six reported are the full candidate set rather than the survivors, and none was
adopted. Finally, the Q1 validation-loss endpoint was changed from the training-log
convention to a lower-variance holdout-shard instrument after data on that axis had
been inspected (Section~\ref{sec:troute}); both readpoints and the instrument change
are disclosed. Each amendment made a claim weaker or a bound looser, which is the direction
an amendment discovered by looking at data usually does not take, and it is the reason
they are declared here rather than beside the results they touch.

A separate change was considered and refused, and it is the more instructive one. An
independent scan re-read $\alpha$ at worlds 8 and 16 after the constants were frozen
and disagreed with the table in opposite directions at the two worlds. Adopting the
world-8 reading alone would have cleared a corpus that currently fails; adopting the
world-16 reading from the same scan would have reddened one that currently passes.
Taking only the half that helps is selecting by outcome, so neither was taken
(Section~\ref{sec:tier1b}). A reporting policy is only worth stating if it constrains
the authors when it is inconvenient, and this is where it did.

\section{The Cost Model}
\label{sec:model}

\subsection{One collective call}

We price a single all-to-all as

\begin{equation}
t(w, S) = \alpha(w) + \frac{S \cdot (w-1)/w}{\beta_{\mathrm{eff}}(S/w)},
\qquad
\beta_{\mathrm{eff}}(x) = \beta_\infty \frac{x}{x + \xhalf},
\label{eq:call}
\end{equation}

where $w$ is the number of ranks in the collective, $S$ is the total send buffer per
rank, $S(w-1)/w$ is the payload that actually crosses a link once the self-copy is
excluded, and $x = S/w$ is the per-peer message size. The fixed term $\alpha(w)$ is
measured per world size. The bandwidth term is the Hockney form~\cite{hockney1994} with a saturating
bandwidth, where $\xhalf$ is the classic half-performance message size.

The saturation term is there because a single flat bandwidth over-credits small
messages, a bias present on every size-sweep corpus and always of the same sign.
Section~\ref{sec:identifiability} sizes that bias and scores this form against the two
other candidates.

\subsection{Strategy costs}

One-hop dispatch is a single call on the full fabric. Two-hop is

\begin{equation}
t_{\mathrm{2hop}} =
\underbrace{t(N_g,\, T M H b)}_{\text{Hop A, slow level}} +
\underbrace{t(R,\, T k H b)}_{\text{Hop B, fast level}} +
t_{\mathrm{splits}} + c_{\mathrm{chain}} \cdot T k ,
\label{eq:twohop}
\end{equation}

with $b$ bytes per element. The last two terms are what makes the accounting
non-obvious. $t_{\mathrm{splits}}$ is the host-side retrieval of split sizes that a
variable-length exchange requires, measured at 0.044~ms per call. The term
$c_{\mathrm{chain}} \cdot Tk$ is the \emph{arrival chain}: the local tensor work at
the representative rank that expands received pairs into the intra-group scatter
plan. We measure it at $0.0875\,\mu$s per row as a PyTorch operator chain. It is
linear in rows, not constant, and at realistic geometries it is the single largest
term in the two-hop overhead.

Equations~\ref{eq:call} and~\ref{eq:twohop} are mean-load expressions. They assume
uniform destination mass when converting a total send buffer into the per-peer size
used by $\beta_{\mathrm{eff}}$. The routing constraints bound each token's fan-out and
quota, but they do not make the aggregate traffic matrix uniform. If the busiest peer
carries $\rho_{\max}$ times the mean load, the relevant message size and completion
time can be inflated by that factor; the released code exposes this as a separate skew
sensitivity rather than fitting it into the balanced microbenchmark. Consequently,
the base model is a balanced-load estimate, not a tail-latency guarantee.

Pricing the two calls of Equation~\ref{eq:twohop} serially is a measured fact rather
than an assumption: per-call cost stops falling at $N \approx 16$ back-to-back
collectives and is flat from there to $N = 1024$, so two collectives cost twice one
(Section~\ref{sec:launch}).

\subsection{Calibrated constants}
\label{sec:constants}

Every constant in Equation~\ref{eq:call} points at a measurement.
Table~\ref{tab:constants} lists them for machine A together with how
each was obtained, because the provenance determines what may be borrowed across
machines and what may not.

\begin{table}[t]
\centering
\small
\begin{tabular}{llp{7.2cm}}
\toprule
Constant & Value & Provenance and transferability \\
\midrule
$\alpha(8), \alpha(16), \alpha(128)$ & 0.111, 0.157, 0.378~ms &
Direct measurement at one token per rank. A machine property, not transferable: on
this machine $\alpha(16) + \alpha(8) \approx \alpha(128)$, so two hops save almost
nothing on fixed cost, while a second machine shows a different shape. Two independent
measurements of this machine disagree, an in-session call-count scan
(Section~\ref{sec:launch}) reading 129 and 134~$\mu$s at worlds 8 and 16 against the
tabulated 111 and 157, a 16\% and a 15\% disagreement in opposite directions. Both sit
inside documented drift and neither is adoptable alone without selecting by outcome
(Section~\ref{sec:tier1b}). \\
$\alpha(256), \alpha(512)$ & 0.735, 1.859~ms &
\textbf{Low confidence.} The only corpus reaching these worlds is the one the model
fits worst, at 40\% median against 5 to 11\% elsewhere, and its absolute bandwidth sits
five times below every other. Results are stated only to world 128
(Section~\ref{sec:coverage}). \\
$\beta_\infty$ & 117.8~GB/s &
Fitted on this machine's own sweeps with $\alpha$ pinned. Independent fits across
four corpora on two machines give 111.0 to 130.3, a spread narrower than either
machine's run-to-run drift. \\
$\xhalf$ & 54~KiB, 90\% CI [30, 87] &
A \emph{borrowed shape}: this machine's corpus has six distinct sizes and cannot
resolve the parameter, so it comes from a 19-size sweep on the second machine. The
entire confidence interval passes the Tier-1 gate. That sweep is corpus C3
(Section~\ref{sec:corpora}), which also supplies 44 of the Tier-1b targets. \\
$\beta_{\mathrm{fast}}$ & 122.4~GB/s (flat) &
Physics-endorsed rather than fitted: the aggregate egress of the intra-node links
predicts 122.4 and measurement gives 122.6, a 0.2\% agreement, with the steadiest
run-to-run spread in the dataset. \\
$c_{\mathrm{chain}}$ & 0.0875~$\mu$s/row & Measured on the operator chain at $H = 2048$.
An implementation property: it transfers across machines running the same software and
not otherwise. Swept over nine row counts and four hidden widths on two nodes, and
verified linear in rows only above 8192 rows and only at fixed $H$: at 1024 rows the
measured chain is 2.8 times the linear prediction, and the $H$ dependence adds
0.203~ms per 1024 of hidden width. The tabulated value is therefore a lower bound
below 8192 rows and above $H = 2048$ (Section~\ref{sec:chainscaling}). \\
\bottomrule
\end{tabular}
\caption{Calibrated constants and what each is. The distinction between machine
properties, physics, and implementation properties is what determines whether a
number may be carried to a new machine or must be re-measured there.}
\label{tab:constants}
\end{table}

\subsection{Model form selection and parameter identifiability}
\label{sec:identifiability}

Two parameters of Equation~\ref{eq:call}, the fixed cost $\alpha$ and the
half-performance size $\xhalf$, are not jointly identified by a size sweep. A
round-trip on synthetic data reproduces predictions to within $10^{-11}$ relative
error while recovering $\xhalf$ 18\% away from truth, so the fit residual carries no
information about which of the two is wrong. \emph{A model form cannot be judged while
one of its parameters is unidentified}, and free fitting therefore does not test the
form at all.

The protocol follows from that property. We measure $\alpha(w)$ directly, pin it, and
fit only the bandwidth pair, resolving $\xhalf$ on C3, the one corpus with enough
distinct sizes to constrain it. That gives 54~KiB rather than the 320~KiB a free fit
returns, and at that value the same model form halves the error instead of tripling
it.

Three candidate forms were scored against C1 to C3 under that protocol and rejected. A
single flat bandwidth runs 8\% to 27\% fast on every size-sweep corpus with the same
sign throughout. The saturating form with $\xhalf$ fitted freely gives 17.9\% median
and 101.6\% worst-case, against 8.1\% and 12.5\% with $\alpha$ pinned. Replacing the
additive Hockney form with $\max(\alpha, \text{wire}/\beta)$, on the theory that a
large call's data movement overlaps the next call's fixed cost, is worse on every one
of the four corpora, at 17\% to 36\% median against 2\% to 15\%.

The 17.9\% median of the freely fitted saturating form was at first taken as evidence
against the model structure when it is a property of the estimator. The consequence
for what follows is that Tier-1 scores the model on the corpus its form was selected
against, which makes Tier-1 a fit-quality check rather than independent evidence for
the form. C4 is the only corpus in this paper that the selection did not touch.

\subsection{Decomposition of the fixed cost}
\label{sec:launch}

Our $\alpha(w)$ lumps the cost of getting a collective onto the device queue together
with the collective's own fixed cost, which are the host overhead and the network
latency that LogP~\cite{culler1993logp} keeps apart. Two-hop issues one more call than one-hop, so
the two parts behave differently under it, and corpus A1 (Section~\ref{sec:corpora})
measures them apart by scanning the same back-to-back collectives in two host
conditions, one with the host running ahead of the queue and one with it waiting for
each call to land before issuing the next. The fixed costs that scan reads disagree
with the $\alpha$ row of Table~\ref{tab:constants}.

\begin{figure}[t]
\centering
\includegraphics[width=\textwidth]{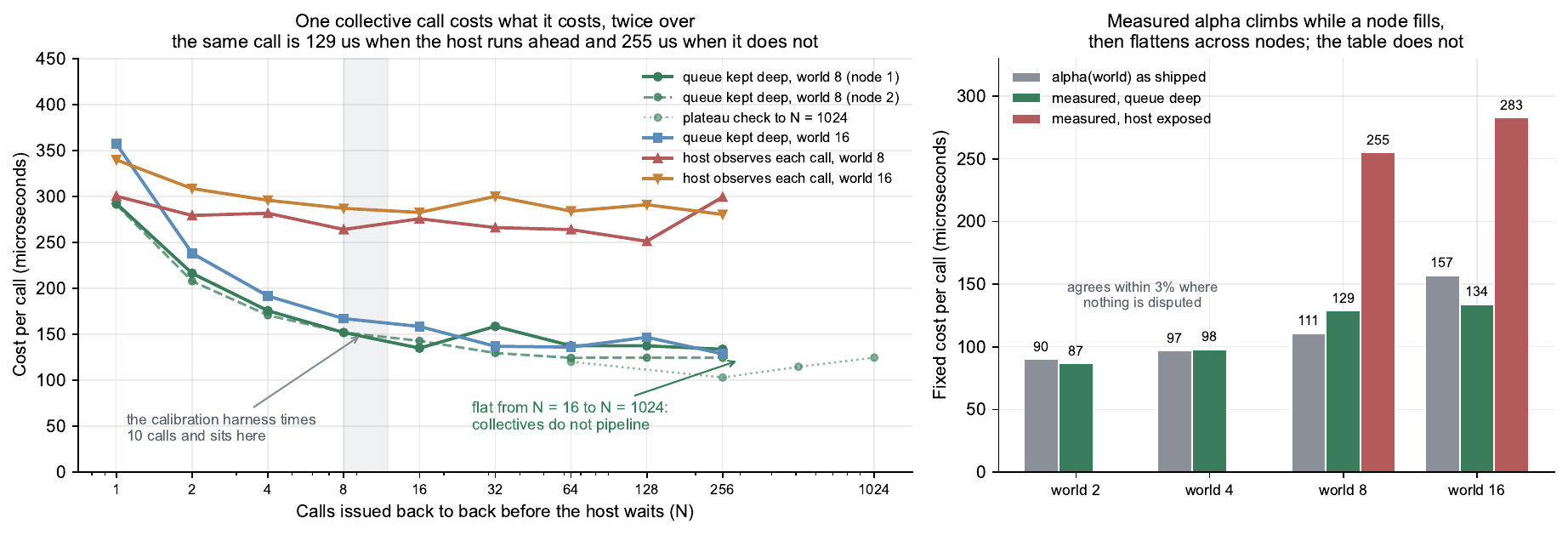}
\caption{Per-call cost of one collective. Left: cost per call against how many calls
are issued before the host waits. It stops falling at $N \approx 16$ and holds flat
to $N = 1024$, so collectives do not pipeline; the upper curves are the same calls
with the host observing each one. Right: the measured fixed cost against the
tabulated $\alpha$, at four worlds.}
\label{fig:launch}
\end{figure}

Figure~\ref{fig:launch} plots the scan. Per-call cost stops falling at $N$ around 16
and is flat from there to $N = 1024$, so
collectives do not pipeline and two back-to-back collectives cost twice one, which is
the assumption behind pricing the two-hop chain serially. From 256~B to 16~KiB the
per-call number is unchanged within the run-to-run spread, as it must be, since the
wire term at 16~KiB is 0.12~$\mu$s. That flat per-call cost is half host cost, and
the host half is hidden only while the queue is deep.
With the host running ahead one call costs 129~$\mu$s at world 8 and 134 at world 16,
and those are the readings that depart from the 111 and 157 of
Table~\ref{tab:constants}; with the host observing each call the same call costs 255
and 283, and the gap holds between 39\% and 59\% of the total at every payload from
64~KiB to 16~MB, so it is a per-call cost that more traffic does not amortise.

\section{Validation Results}
\label{sec:validation}

Six corpus outcomes follow, four passing and two failing, each scored against the
estimand, the threshold and the registration status stated in
Section~\ref{sec:gatereg} and collected in Table~\ref{tab:gateresults}.

\begin{table}[t]
\centering
\small
\begin{tabular}{@{}llllc@{}}
\toprule
Gate & Corpus & Statistic & Signed bias & Outcome \\
\midrule
Tier-1  & C1 & 4.1\% median, 24.5\% worst, crossover in window & --- & \textbf{pass} \\
Tier-1b & C2 & 1.9\% median  & $+0.6\%$ & \textbf{pass} \\
Tier-1b & C3 & 9.3\% median  & $-1.1\%$ & \textbf{pass} \\
Tier-1b & C4 & 8.0\% median  & $-2.0\%$ & \textbf{pass} \\
Tier-1b & C5 & 15.1\% median & $-9.1\%$ & \textbf{fail} \\
Tier-2  & G1 & MAE 0.135, 0/6 in band & --- & \textbf{fail} \\
\bottomrule
\end{tabular}
\caption{Gate outcomes. Corpus labels, estimands, thresholds, registration status and
the capability each failure removes are in Table~\ref{tab:gatereg}. C2 to C4 are
scored as a conjunction; C5 is scored separately. The two failures are analysed in
Sections~\ref{sec:tier1b} and~\ref{sec:tier2}.}
\label{tab:gateresults}
\end{table}

\subsection{Tier-1: communication micro level}
\label{sec:tier1}

Tier-1 passes, and what makes the pass mean anything is how its targets were built
rather than the size of the margin. Targets are pooled medians over repeated runs, not
single runs, because the instrument drifts by more than the effect
(Section~\ref{sec:protocol}), and an earlier version of this gate passed on single-run
targets that a different run would have failed. Tier-1 also scores the model on the
corpus its form was selected against (Section~\ref{sec:identifiability}), which is why
C4 and not C1 is the corpus that tests prediction.

\subsection{Tier-1b: an independent benchmark family, on two machines}
\label{sec:tier1b}

Tier-1 is anchored to the benchmark family the calibration came from, which makes it
internally consistent and self-referential. Tier-1b scores the same model against
plain all-to-all size sweeps collected weeks apart by a different benchmark, on two
machines. On machine B, $\alpha$ and flat bandwidth are refitted, while the
$\xhalf$ shape and the level/arrival-chain assumptions are retained.

C3 supplies both the size sweep that resolves $\xhalf$ and the 44 targets scored
here, so for that parameter it is a consistency check rather than a prediction, and
the licensing role falls to C4 instead. C3 is therefore a same-corpus fit/consistency
check: the functional form retains acceptable error after machine-specific refitting,
but it is not an out-of-sample transfer test.
C4 is the only corpus collected after the constants were frozen with nothing fitted to
it, so it is the only one in this paper that tests prediction rather than consistency,
and it is the only Tier-1b row that is an out-of-sample prediction rather than a
regression guard (Section~\ref{sec:gatereg}).

C5, a world-8 sweep on machine A of thirteen sizes and twelve repetitions each,
misses the Tier-1b gate at 15.1\% median with a $-9.1\%$ bias. The miss is localized
rather than diffuse: below 8~MB, where $\alpha$ is 76 to 97\% of the prediction, the
model runs 18\% fast; above it, where the wire term dominates, it runs 10\% slow. That
is one number, $\alpha(8) = 0.111$, against 0.129 measured the same day by the call-count
scan of Section~\ref{sec:launch}. Setting $\alpha(8)$ to the measured value clears
the corpus at 9.3\% and leaves Tier-1 at 4.9\% against its 20\% threshold.

We did not make that change, for a reason that is not conservatism. The same scan
reproduces $\alpha$ at worlds 2 and 4 to within 3\%, so the instrument is sound, and
it also reads $\alpha(16)$ at 134~$\mu$s against the tabulated 157. Adopting the
world-8 reading turns a red corpus green; adopting the world-16 reading from the same
scan turns a green corpus red, from 8.0\% to 15.5\%. Taking only the half that helps
would be selecting by outcome. Both readings sit inside the documented run-to-run
drift (Section~\ref{sec:protocol}), and the choice moves the effective breakeven by
under 2.1\%,
so nothing in this paper depends on it. The corpus is reported failing, with its
cause, and a test pins both halves of the scan: that the world-8 reading would clear
this corpus, and that the world-16 reading from the same scan would turn a passing one
red. With both halves pinned, the choice cannot later be revised on one side only.

\subsection{Tier-2: end-to-end step time}
\label{sec:tier2}

Tier-2 fails, and the consequence registered in Section~\ref{sec:gatereg} takes
effect: the model refuses to produce a step-level prediction, and this paper makes
none anywhere.

The cause is nameable. Summed phase-level timings and the step-level delta disagree
by roughly a factor of five, because the two arms overlap communication with compute
differently and an event-timed phase span says nothing about what ran beside it.

\begin{figure}[t]
\centering
\includegraphics[width=0.95\textwidth]{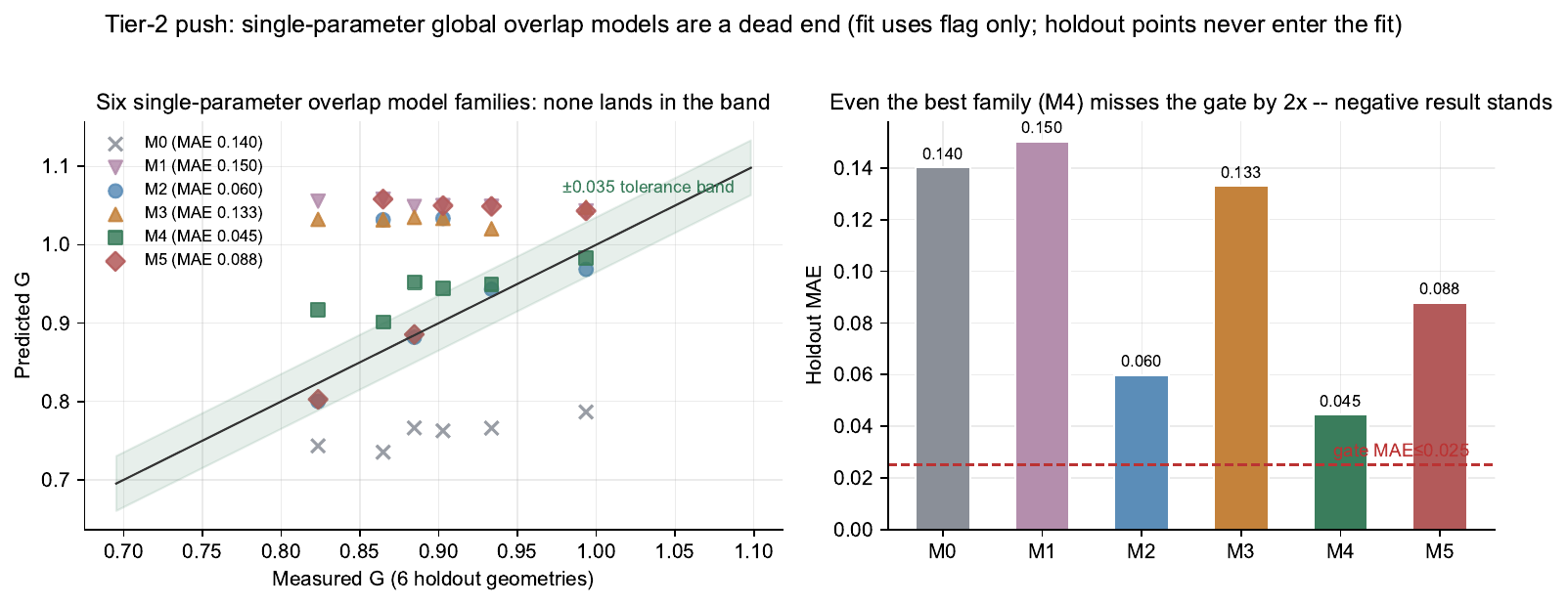}
\caption{Predicted against measured step-time ratio for the six overlap families.
The shaded band is the $\pm 0.035$ tolerance the gate requires four of six holdouts
to fall inside. No family places a single point in it.}
\label{fig:overlap}
\end{figure}

A single overlap parameter does not close that gap. The six families
below are exploratory in the sense declared in Section~\ref{sec:policy}: they were
enumerated after Tier-2 failed, in order to unlock it, with no prespecified candidate
set and no stopping rule, and none was adopted. Fitted on one calibration geometry and
evaluated on the six holdouts of G1, they range from no overlap at all to exposure
proportional to pipeline pressure (Table~\ref{tab:overlap},
Figure~\ref{fig:overlap}). \textbf{None passes.} The closest in magnitude, hiding per
call at MAE 0.060, predicts the wrong sign on the scale axis. The one with all signs
correct, hiding proportional to compute, misses the error gate by nearly a factor of two at
MAE 0.045 against a threshold of 0.025.

\begin{table}[t]
\centering
\small
\begin{tabular}{llccc}
\toprule
Family & Structure of the exposed step delta & Fitted parameter & Holdout MAE & Signs \\
\midrule
M0 & $\Delta = \Delta_{\mathrm{model}}$ (no overlap) & --- & 0.140 & 5/5 \\
M1 & $\varphi \, \Delta_{\mathrm{model}}$ & $\varphi = -0.153$ & 0.150 & 0/5 \\
M2 & $\Delta_{\mathrm{model}} - h \cdot \mathrm{calls}$ & $h = 2.51$~ms & \textbf{0.060} & 3/5 \\
M3 & $\Delta_{\mathrm{model}} - \varphi \cdot \mathrm{fixed}$ & $\varphi = 1.14$ & 0.133 & 0/5 \\
M4 & $\Delta_{\mathrm{model}} - c \cdot T_{\mathrm{comp}}$ & $c = 0.299$ & \textbf{0.045} & \textbf{5/5} \\
M5 & $\Delta_{\mathrm{model}} (1 - \lambda / \mathrm{mbs})$ & $\lambda = 1.153$ & 0.088 & 2/5 \\
\bottomrule
\end{tabular}
\caption{Six single-parameter overlap families, each fitted on the one calibration
geometry of G1 and scored on its six holdouts against the Tier-2 gate of
MAE $\leq 0.025$. MAE is over the holdouts; the sign column is over the geometries
where a sign is defined (Section~\ref{sec:gatereg}). M0 is not the Tier-2 row of
Table~\ref{tab:gateresults}, which reads MAE 0.135 on the same six holdouts: that row
back-solves one combine constant on the calibration geometry, while M0 fits nothing.}
\label{tab:overlap}
\end{table}

The obstruction is structural rather than a matter of tuning. One calibration point
determines exactly one parameter, and using holdouts to fit more would cancel the
validation. More importantly, the implied exposure ratio in the data tracks
micro-batch count, $-0.15$, $+0.42$, $+0.62$ for one, two and four micro-batches,
which says exposure is a property of the \emph{schedule} and not of the model. What
is missing is a different measurement rather than a better fit. Exposure has to be
attributed per stream, as span minus the intersection with compute-stream busy time,
and the two arms' exposure difference has to reconcile with their step-time difference
before any overlap parameter is identifiable at all. No instrument we ran resolves it
at that granularity, and the tier stays locked.

\subsection{Calibration uncertainty}
\label{sec:uncertainty}

\begin{figure}[t]
\centering
\includegraphics[width=0.86\textwidth]{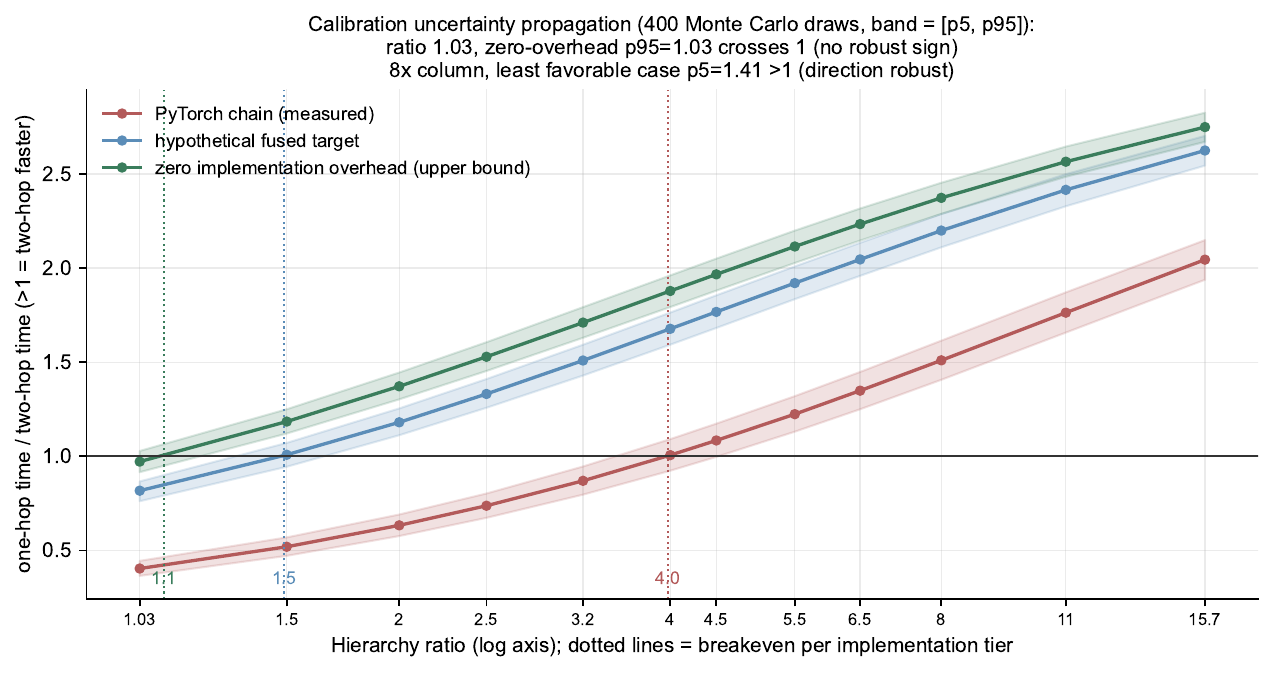}
\caption{Calibration uncertainty propagated through the extrapolation, 400 Monte Carlo
draws over the measured spread of every constant including the bootstrap interval of
$\xhalf$. Bands are [p5, p95].}
\label{fig:uncertainty}
\end{figure}

Propagating the measured spread of every constant through the extrapolation, over 400
Monte Carlo draws including the bootstrap interval of $\xhalf$, bounds how much of a
verdict is calibration error (Figure~\ref{fig:uncertainty}). At hierarchy ratio 8 the
least favourable draw is still 1.41, so the direction of the
communication-call sensitivity is robust to the propagated calibration error. At
ratio 1.03 the zero-overhead p95 reaches 1.03, crossing unity; no robust sign claim is
made for that corner.

The gates constrain $\xhalf$ from above at 77~KiB and not at all from below. An
earlier sweep of $\xhalf$ against every gate gave a two-sided admissible range of
[46, 76]~KiB, half the width of the bootstrap interval; six repetitions of the corpus
that produced its lower bound, scored at the median, dissolved that bound, and the
interval is withdrawn. The error ran in the flattering direction, which is the part
worth recording. \emph{An admissibility interval computed from a noisy corpus comes
out too tight}, and too tight reads as a stronger result than the data supports.

\subsection{Coverage limits above 128 ranks}
\label{sec:coverage}

One corpus is the sole source of every number above world 128, and it is the corpus
this model fits worst. That one fact costs this paper a claim.

\begin{figure}[t]
\centering
\includegraphics[width=0.86\textwidth]{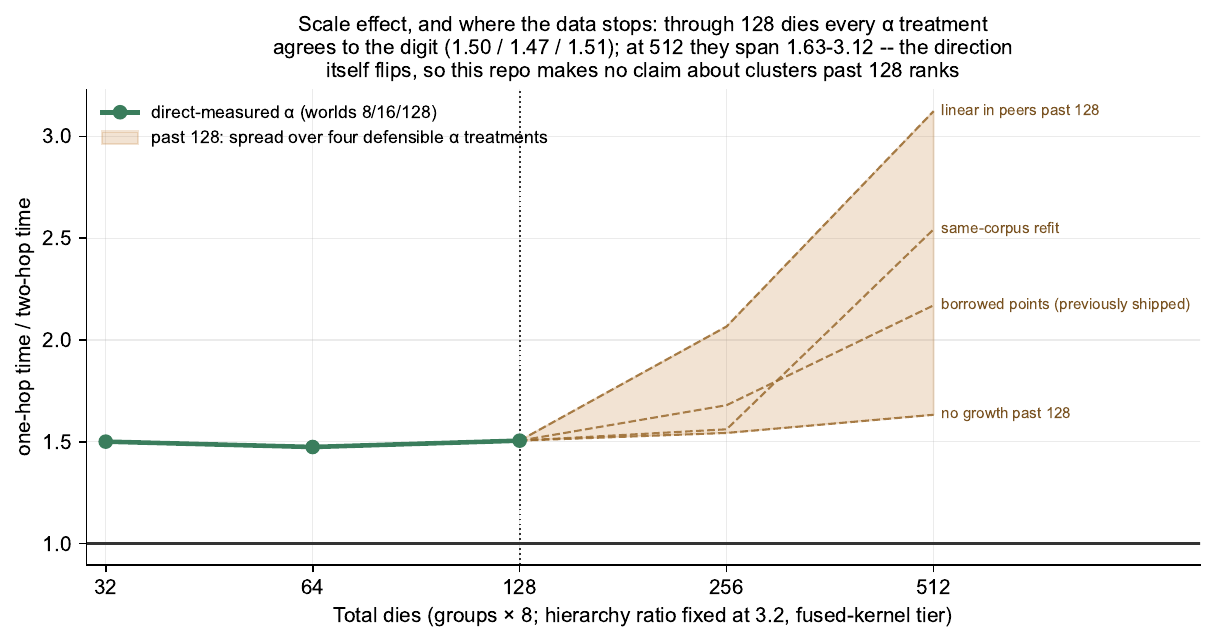}
\caption{Scale extrapolation. Through 128 ranks every defensible treatment of
$\alpha$ agrees to the digit. Past 128 the figure plots the band across four
treatments rather than a line, because the only corpus covering those worlds is the
one we have quantitative reason to distrust.}
\label{fig:scale}
\end{figure}

An earlier version of this work reported that two-hop's advantage grows with cluster
size, quoting a ratio of 1.94 at 512 ranks and attributing the gain to $\alpha(w)$
growth. An audit of all 331 usable size-sweep points establishes that the only corpus
reaching worlds 256 and 512 is also the corpus whose absolute bandwidth sits five
times below every other and which the model fits worst, at 40\% median relative error
against 5 to 11\% elsewhere.

Propagating four defensible treatments of those two $\alpha$ entries leaves the
512-rank ratio anywhere between 1.37 and 2.94, and the direction of the trend flips
between treatments. Figure~\ref{fig:scale} plots that band, and we therefore withdraw
the claim. The paper states results for worlds up to 128, where every treatment
agrees. Both halves of that boundary are enforced by a check, the lower half so the
coverage is not understated and the upper half so the claim cannot be quietly
re-hardened (Section~\ref{sec:repro}).

A world-scaling bandwidth degradation term appears in one corpus at matched per-peer
size, roughly $0.55\times$ at 256 ranks and $0.36\times$ at 512, but that is the same
low-confidence corpus, so the term is recorded as unconfirmed and left out.

\section{Applicability}
\label{sec:applicability}

With the model admitted at the communication-call level and locked at the step level,
the defensible output is a sensitivity study. A target deployment requires
its own fast/slow collective measurements and a passing communication-level gate;
nominal link ratios alone do not license a prediction.

\subsection{Breakeven}
\label{sec:breakeven}

Counting only bytes, two-hop wins when

\begin{equation}
\rbe = \frac{(1 - 1/R)\, q}{q - 1} < \frac{\bfast}{\bslow},
\label{eq:breakeven}
\end{equation}

which at $R = 8$ gives 1.75 for $q=2$, 1.31 for $q=3$, and 1.00 for $q=8$. In this
expression $q$ appears on both sides of the trade, and the group size $R$ enters only
through the share of fast-side traffic that stays local.

Equation~\ref{eq:breakeven} is not the operative threshold, because it ignores the
fixed cost of the extra collective and the arrival chain. Running the full model to
the point where the two strategies tie gives an \emph{effective} breakeven of
3.98 with the measured PyTorch operator chain, 1.49 for a hypothetical fused target,
and 1.10 at zero implementation overhead. The gap between 1.31 and 3.98 is the price of the
implementation. It is larger than the difference between many pairs of real
interconnects, but this model has not been calibrated on a machine in that interval. The
DeepSeek-V3 training node reports NVLink at 200~GB/s against 50~GB/s per InfiniBand
NIC, and 160 against 40 once achieved rather than nominal figures are used, which is a
hierarchy ratio of 4.0 on either accounting~\cite{deepseekv3hw}. That node sits just
above the operator-chain threshold of 3.98 in the ratio-only calculation and far above
the hypothetical fused threshold. This comparison illustrates sensitivity; it is not
a prediction for that node because the other machine constants were not measured
there. Quoting the achieved bandwidth on the fast side against the
nominal figure on the slow side, which is the natural way to read that sentence, would
have put it at 3.2 and on the wrong side of the threshold; the two sides of a ratio
have to be measured the same way. All three figures are stated at the reference operating point
used throughout this paper, $H = 2048$, 24576 rows, $R = 8$, $q = 3$, worlds up to 128
(Section~\ref{sec:corpora}). Section~\ref{sec:chainscaling} shows the threshold is not
flat in $H$, and every later mention below is a displacement of these figures rather
than a fresh calibration.

\subsection{Sensitivity to hidden width}
\label{sec:chainscaling}

The effective breakeven of Section~\ref{sec:breakeven} is stated at $H = 2048$, which
is where $c_{\mathrm{chain}}$ was calibrated, and the arrival chain is not independent
of $H$. Sweeping the operator chain over hidden widths at a fixed 24576 rows, thirty
iterations on each of two nodes, gives 2.37, 2.51, 2.85 and 3.79~ms at $H$ of 1024, 2048, 4096 and 8192. That
decomposes as 2.17~ms of index and sort work plus 0.203~ms per 1024 of hidden width,
so at $H = 2048$ the $[\text{rows}, H]$ gather is only 14\% of the chain and the
index work is the other 86\%.

Propagating those measurements requires moving $H$ in both places at once, because
the payload the chain is weighed against widens with $H$ as well. Over the fourfold
range from 2048 to 8192 the bytes each collective carries rise by four and the measured
chain by 1.51, so the chain's share of two-hop falls from 46\% to 27\% and two-hop
improves as hidden width grows. The effective breakeven therefore \emph{falls} with
$H$: 5.96 at 1024, 3.98 at the reference width, 2.90 at 4096 and 2.40 at 8192. An
earlier version of this section scaled the chain with $H$ while holding the payload at
the reference width and reported 4.91 and 6.17, which has the sign of the effect
backwards. The limit settles the direction without reference to any constant: every
wire term scales exactly with $H$ while $\alpha$ and the splits exchange do not, so at
large $H$ the comparison approaches the byte-only one and the threshold must fall
toward it rather than away.

Contemporary MoE models sit above the reference width, with $H = 7168$ in DeepSeek-V3,
5120 in Pangu Pro MoE and 4096 in the 352B production model of
MegaScale-MoE~\cite{deepseekv3,tang2025pangumoge,jin2026megascalemoe}. A threshold
stated at $H = 2048$ and applied there therefore \emph{over-prices the arrival chain,
and the arrival chain is paid only by two-hop}. That bias runs against the method this
paper proposes rather than in its favour, and it is stated here because the previous
version of this section claimed the opposite and offered the claim as a disclosure. The
reference figure of 3.98 is the conservative end of the range reported here and not the
flattering one.

One convention runs through every point above: the arrival-chain level of
Section~\ref{sec:constants}, 2.15~ms at 24576 rows from a single measurement, carrying
the shape the hidden-width sweep measures. The sweep's own $H = 2048$ point is 2.51~ms,
from eighteen measurements, and adopting that level instead would put the reference
threshold at 4.46. Both are reported because the gap between them is the run-to-run
drift already documented for that constant, and because the choice moves the level of
every figure in this subsection without moving the direction.

The chain is dominated by
index work rather than by data movement, which is precisely what a fused kernel
removes, so the case for fusing it is stronger at large $H$, not weaker. The
linear-in-rows form we use is verified only above 8192 rows: at 1024 rows the chain
measures 0.248~ms against the 0.090~ms the linear form predicts, so small geometries
carry the same optimistic bias. Neither conclusion depends on the calibration.

\subsection{Where the criterion's sensitivity lives}
\label{sec:sensitivity}

The two implementation overheads the model prices are not comparable. Charging two-hop
one extra host exposure, which Section~\ref{sec:launch} measures at 39\% to 59\% of a
call's cost, moves the effective breakeven of Section~\ref{sec:breakeven} to
\textbf{4.15}, from 3.98. Removing the arrival chain instead moves it to
\textbf{1.10}. The two shifts are 0.17 and 2.88 of hierarchy ratio. Thus the
arrival-chain term dominates this local sensitivity calculation. This ordering does
not establish that implementation decides the sign on an unmeasured target machine;
it identifies the first implementation term to remeasure or optimize.

\subsection{Ratio-only scenarios, not target-platform predictions}
\label{sec:platform}

Table~\ref{tab:predict} holds platform A's calibrated constants and the reference
geometry fixed, then changes only the slow-side bandwidth to realize several hierarchy
ratios. It is a scenario/sensitivity table. Platform B is absent because its corpus
does not separate fast and slow levels, and naming a commercial platform here would
incorrectly imply that its remaining constants had been measured.

\begin{table}[t]
\centering
\footnotesize
\begin{tabular}{@{}p{3.9cm} r r r r r@{}}
\toprule
Scenario & Ratio & Measured chain & Hypothetical fused & Zero overhead
& Chain pays above \\
\midrule
platform A, measured constants & 1.03 & 0.37 & 0.80 & 0.98 & no width \\
synthetic ratio sensitivity & 2.0 & 0.63 & 1.18 & 1.37 & beyond 16384 \\
synthetic ratio sensitivity & 4.0 & 1.00 & 1.67 & 1.87 & 2000 \\
synthetic ratio sensitivity & 9.0 & 1.60 & 2.28 & 2.44 & 490 \\
synthetic ratio sensitivity & 18.0 & 2.14 & 2.69 & 2.80 & 260 \\
\bottomrule
\end{tabular}
\caption{Ratio-only sensitivity. The three middle columns are one-hop over two-hop
\emph{communication-call} time at the reference geometry, above 1 favouring two-hop in
the model. The last column is the hidden width above which the measured chain pays at
that ratio, obtained by bisection on $H$ with the chain carrying the shape measured in
Section~\ref{sec:chainscaling} and the payload widening with $H$; it is the same
calculation as the three columns beside it, read along the axis those columns hold
fixed. The reference width of 2048 is the least favourable one for two-hop, so the
middle columns understate every row: platform A rises from 0.37 to 0.63 as $H$ grows
without ever reaching 1, while ratio 4 crosses at almost exactly the reference width.
The two extreme entries in the last column are extrapolations of the chain fit well
outside the measured range of 1024 to 8192 and are given to one significant figure for
that reason. Only the first row uses a measured ratio; the other rows change platform
A's slow-side bandwidth synthetically. The 0.012~$\mu$s/row fused column is a
hypothetical target, not a measured kernel. No row is a target-platform or
training-throughput prediction, and balanced routing is assumed.}
\label{tab:predict}
\end{table}

The table shows three things and no more. First, the measured chain crosses near ratio
4 at the reference width, while the hypothetical target crosses below 2; implementation
cost materially changes the communication-call sensitivity. Second, changing
provisioning can move a deployment horizontally even when accelerator and link
technologies are unchanged. Third, the ratio axis is not the only one that moves a row:
the last column reads the same model along hidden width, and a machine at ratio 9 pays
for any model wider than about 490 while a machine at ratio 2 pays at no width anyone
builds. A row scored at the reference width alone is scored where two-hop does worst.
To turn either observation into a deployment statement, the adopter must measure
$\bfast$, $\bslow$, $\alpha$, $\xhalf$, and the arrival chain at the actual message
sizes, verify the domain-uniformity precondition, and pass the communication-level
gate. The step-level gate must additionally pass before any throughput claim.

\begin{figure}[t]
\centering
\includegraphics[width=0.95\textwidth]{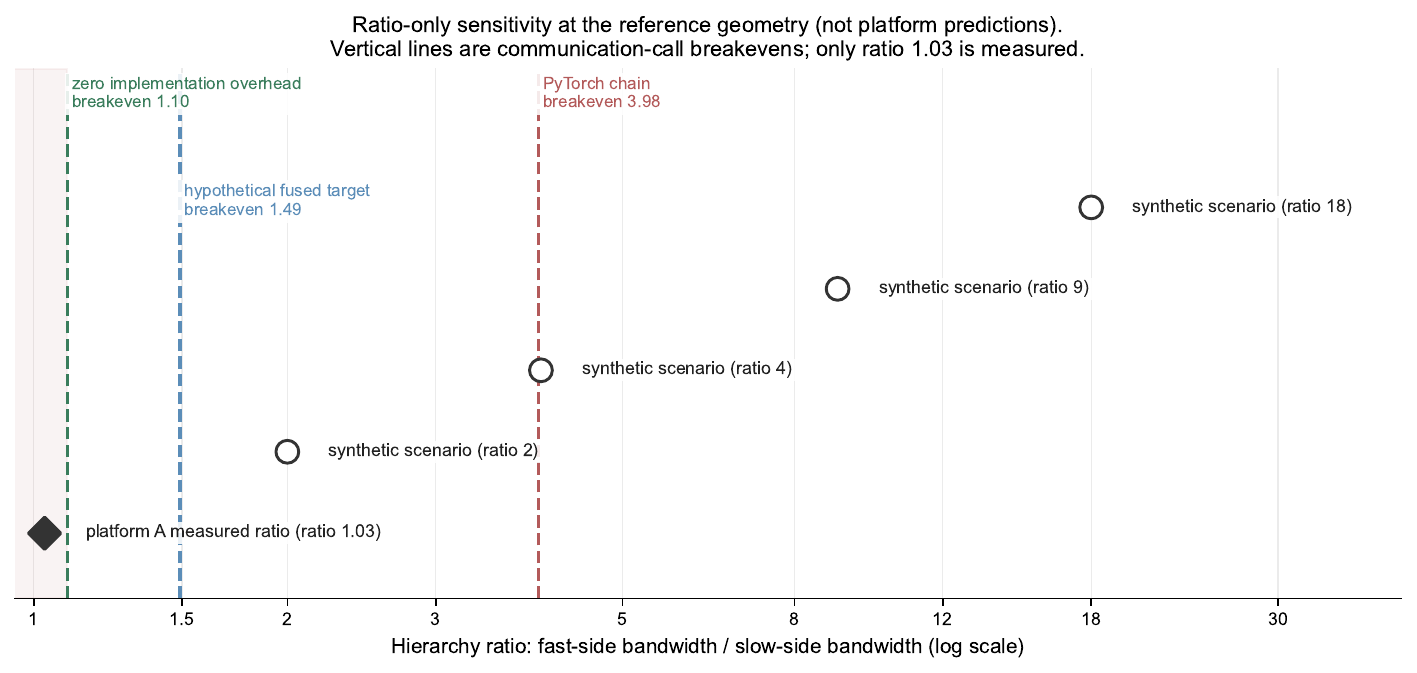}
\caption{Ratio-only sensitivity map. Vertical lines are the effective breakeven
thresholds at the reference geometry: 3.98 for the measured chain, 1.49 for the
hypothetical fused target, and 1.10 at zero overhead. Points above 1.03 are synthetic
ratio scenarios with platform A's remaining constants held fixed; they are not
commercial-platform predictions. Hidden-width measurements move the measured-chain
threshold down, to 2.90 at $H=4096$ and 2.40 at $H=8192$
(Section~\ref{sec:chainscaling}).}
\label{fig:platformmap}
\end{figure}

\subsection{An alternative that was priced and rejected}
\label{sec:onesided}

Kernel-initiated one-sided writes are the obvious candidate for removing collective
overhead on the fast path, obvious enough that a device-initiated library already
builds MoE dispatch on them~\cite{deepep2025}. Measured against collective all-to-all
inside one 8-die node, across two implementations and two core counts, the best
configuration reached $0.68\times$ against a prespecified criterion of
$\geq 1.15\times$ on at least two valid size tiers, and no tier qualified. The upside
is bounded above independently of measurement: the collective already runs at roughly
85\% of the single-die aggregate egress, so the entire available headroom is about
15\%, and the existing one-sided kernel relays every byte through on-chip buffers
twice.

\subsection{A five-condition machine profile and one lever}
\label{sec:profile}

Five conditions have to hold together, and they are not five independent hurdles.
They fall into three stages in sequence, and a machine can fail at any of them for
reasons that look nothing alike.

The first two conditions decide whether there is anything to deduplicate, and both are
properties of the geometry rather than of the fabric. The \emph{fan-out} condition
asks for $q = k/M \geq 2$: at $q = 1$ each token already sends one row per group,
the saving $(q-1)/q$ is zero, and no interconnect can rescue it. The \emph{domain}
condition asks that expert parallelism span more than one fast domain, because if
every expert fits inside one high-bandwidth domain there is no slow-side traffic to
deduplicate at all. This is not merely a negative verdict: whenever the
expert-parallel group fits inside one regular fast domain, the criterion returns that
the cross-domain restructuring does not apply.

The next two decide whether a byte saved becomes time saved, and both are machine
properties that have to be measured rather than assumed. The \emph{saturation}
condition asks that per-peer messages sit comfortably above $\xhalf$, since below the
knee the collective is latency-bound and byte savings do not convert. The
\emph{fixed-cost} condition asks that $\alpha(N_g) + \alpha(R) < \alpha(EP)$: two hops
pay two fixed costs where one paid one, so on a machine whose $\alpha$ barely grows
with world size the swap loses before any byte moves. The machine of Section~\ref{sec:e2e} clears this
condition, at 0.268 against 0.378~ms, but by so little that the saving is immaterial
beside its verdict on bytes, and neither that margin nor the risk it measures is
visible in a topology diagram.

The fifth decides whether the saving exceeds what the restructuring costs. This
\emph{threshold} condition requires hierarchy ratio at or above the \emph{effective}
breakeven for the arrival chain actually in use, not the byte-only figure. It is the only
condition an adopter can move without changing machines, which is why we report it
separately even though it is one line of the same criterion. The arrival chain is
priced into it and is not a sixth condition; separating it out would double-count.
The size of that lever is the point: it is worth 2.88 of hierarchy ratio against the
0.17 the launch path is worth (Section~\ref{sec:sensitivity}).

The profile and the cost model cannot disagree by construction: the threshold
condition uses the effective breakeven computed from the same model that produces the
ratio. We arrived at that construction the hard way. An earlier draft scored the
arrival chain as a separate condition, and it consequently rejected a machine the model
itself scored at 1.65, which is the failure this arrangement exists to prevent. The
chain is therefore priced into the threshold condition rather than listed beside it,
and a test sweeps ratio against implementation tier to enforce that the two can never
diverge again.

\section{Cost of the Enabling Routing Constraint}
\label{sec:troute}

Two constraints are at work and they do different work (Section~\ref{sec:problem}).
Confining each token's experts to $M$ of the $N_g$ groups is what makes two-hop
dispatch expressible. The per-group quota fixes the number of expert rows contributed
by each token to a selected group. It does \emph{not} make the aggregate message to
each peer constant size: group selection and token counts remain data-dependent, so a
counts exchange, padding, or a capacity-limited transport may still be required. The
conjunction therefore gives a compile-time bound on per-token fan-out and multiplicity,
not a static traffic matrix. The routing constraint is priced separately from the
transport so that its measured quality cost can be judged before any two-hop kernel is
built.

\begin{figure}[t]
\centering
\includegraphics[width=0.98\textwidth]{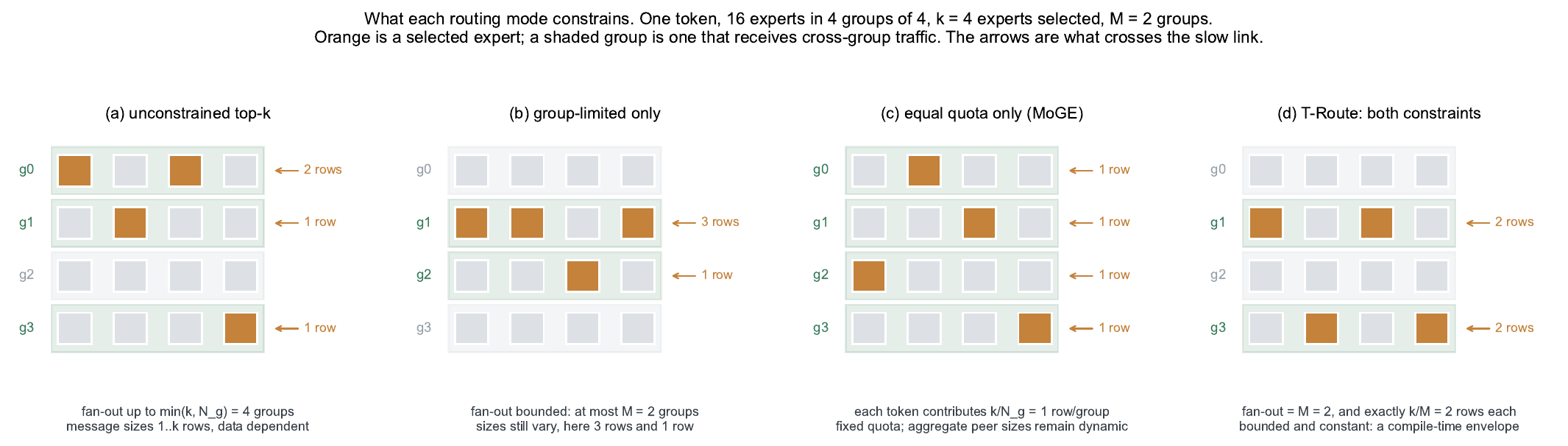}
\caption{What each of the four routing modes constrains. The geometry drawn is
illustrative, 16 experts in 4 groups of 4 with $k = 4$ and $M = 2$, deliberately
smaller than the measured geometry so that all four modes are visibly distinct. Shaded
groups receive cross-group traffic and the annotations give the payload rows crossing
into each. Unconstrained top-$k$ bounds neither how many groups a token reaches nor
the size of the message to each; the group limit alone bounds the count and leaves the
per-token multiplicities data dependent; the equal quota alone fixes a per-token quota
and reaches every group. The conjunction bounds fan-out and per-selected-group
multiplicity, while aggregate peer sizes remain data dependent.}
\label{fig:routemodes}
\end{figure}

The measurement is corpus Q1 (Section~\ref{sec:corpora}): a 13.14B-parameter model
with 1.33B active, $E = 128$ experts in 8 groups of 16, $k = 8$ and $M = 4$, so the
deduplication quota is $q = k/M = 2$, the smallest value at which the dispatch has
anything to deduplicate (Section~\ref{sec:profile}). Four routing modes, unconstrained
top-$k$, group-limited only, equal-quota only, and both
(Figure~\ref{fig:routemodes}), run at four seeds each over 62.9B tokens per arm, giving
sixteen arms. Every mode is selected by a mode argument to a single entry point in
\texttt{terrace/routing.py}, whose whitelist raises on an unrecognised mode rather than
falling through to a default arm; one shared code path is a precondition for the
comparison to mean anything. Endpoints, margins, seeds, tokens per arm and confidence
level were recorded in the campaign protocol; the post-data instrument and checkpoint
changes are disclosed in Section~\ref{sec:policy}.

\begin{table}[t]
\centering
\small
\begin{tabular}{llcc}
\toprule
Mode & Constraint & $\Delta$ val loss vs unconstrained & 90\% CI \\
\midrule
group-limited & group limit only & $+0.00276$ & $[+0.00223, +0.00329]$ \\
quota-only & equal quota only & $+0.00895$ & $[+0.00726, +0.01064]$ \\
\textbf{both} & group limit and equal quota & $\mathbf{+0.00339}$ & $[+0.00224, +0.00455]$ \\
\bottomrule
\end{tabular}
\caption{Quality cost of the routing constraints at the 20k-step readpoint. Holdout
validation loss, paired by seed, against the unconstrained control, with 90\%
confidence intervals.}
\label{tab:quality}
\end{table}

\begin{figure}[t]
\centering
\begin{subfigure}{0.49\textwidth}
\includegraphics[width=\textwidth]{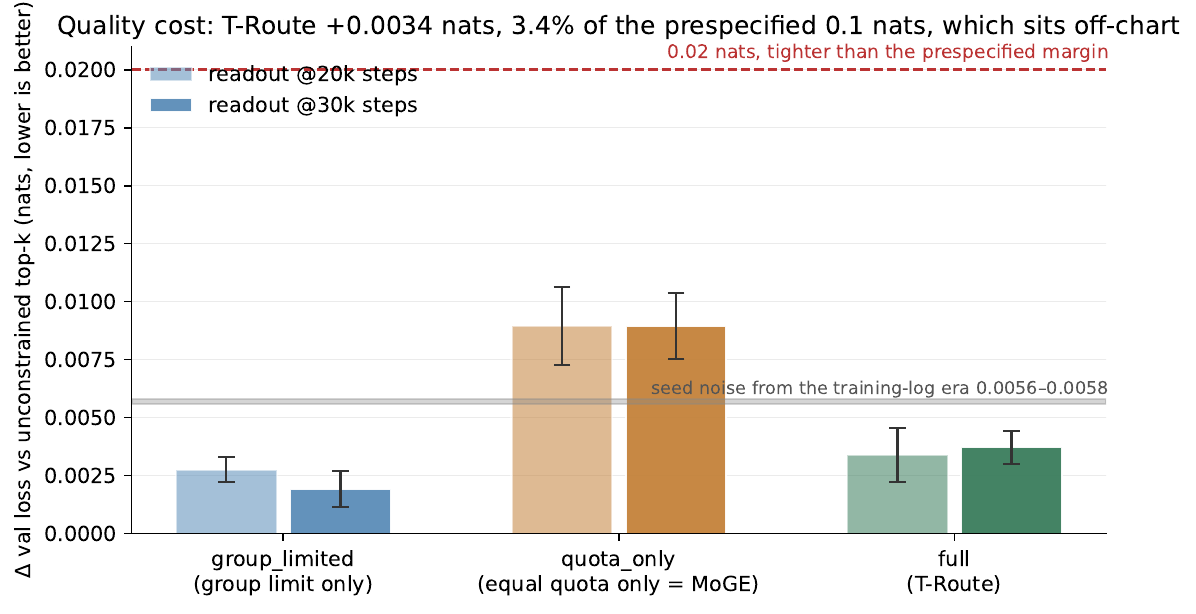}
\caption{Paired deltas with 90\% confidence intervals at two readpoints.}
\end{subfigure}
\hfill
\begin{subfigure}{0.49\textwidth}
\includegraphics[width=\textwidth]{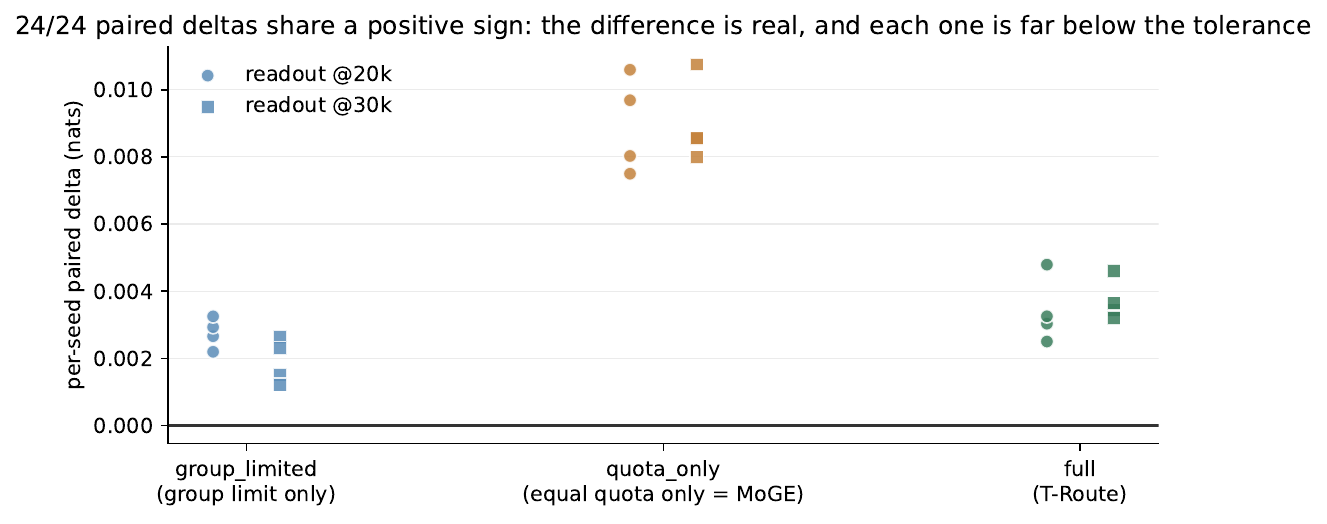}
\caption{The individual paired seed deltas behind those estimates.}
\end{subfigure}
\caption{Quality axis, each constrained mode against the unconstrained control. The
left panel draws a reference line at 0.02 nats, which is tighter than anything
prespecified; the protocol no-loss margin is the 0.1 nats of
Section~\ref{sec:falsification} and lies off the drawn range.}
\label{fig:quality}
\end{figure}

All sixteen arms were read at 20k and at 30k steps on a holdout shard, and
Table~\ref{tab:quality} reports the earlier readpoint. At 30k the three constrained
modes give $+0.00192$ with interval $[+0.00114, +0.00270]$, $+0.00895$ with
$[+0.00751, +0.01039]$, and $+0.00372$ with $[+0.00300, +0.00444]$, so the drift across
the intervening 10k steps is $-0.00084$, zero to five decimals, and $+0.00033$. The
exact repeat on quota-only was checked against the per-seed record rather than taken on
trust: the four per-seed values differ between the readpoints and the means coincide
only once rounded. The conjunction costs $+0.0034$ nats, 3.4\% of the prespecified
no-loss margin of 0.1 nats, and all 24 paired seed deltas are positive, three
constrained modes at four seeds and two readpoints. All three intervals exclude zero,
so the effect is resolved and small rather than absent. That distinction belongs to the
instrument: the holdout probe carries a within-arm standard deviation of 0.0008 to
0.0023 nats, where the training-log convention an earlier analysis used carried
between-seed dispersion of 0.0056 to 0.0058. The change of instrument was made after
data on this axis had been seen and is the fourth amendment logged in
Section~\ref{sec:policy}.

One comparison in the table is not a null, and it is not the one it appears to be.
Equal quota alone costs $+0.00895$ and the conjunction $+0.00339$, with the intervals
disjoint at both readpoints, which invites the reading that adding the group limit to
the quota is free. The two modes are not nested, so that reading does not hold. In
\texttt{terrace/routing.py} the quota-only arm selects $k/N_g = 1$ expert in each of the
eight groups while the conjunction selects $k/M = 2$ in each of four, and in this
geometry their admissible selections are disjoint; what the comparison shows is that
one constraint is cheaper than another. The nesting that does exist runs the other way.
The conjunction and the group-limited arm score and choose groups by the identical rule
and the conjunction then restricts the selection inside those groups, so its admissible
set is a strict subset of the group-limited arm's. Against that reference the quota
costs $+0.00063$ more at 20k and $+0.00180$ more at 30k, the ordinary direction, and
that increment is the figure an adopter who already holds the fan-out bound needs.

\begin{figure}[t]
\centering
\includegraphics[width=0.98\textwidth]{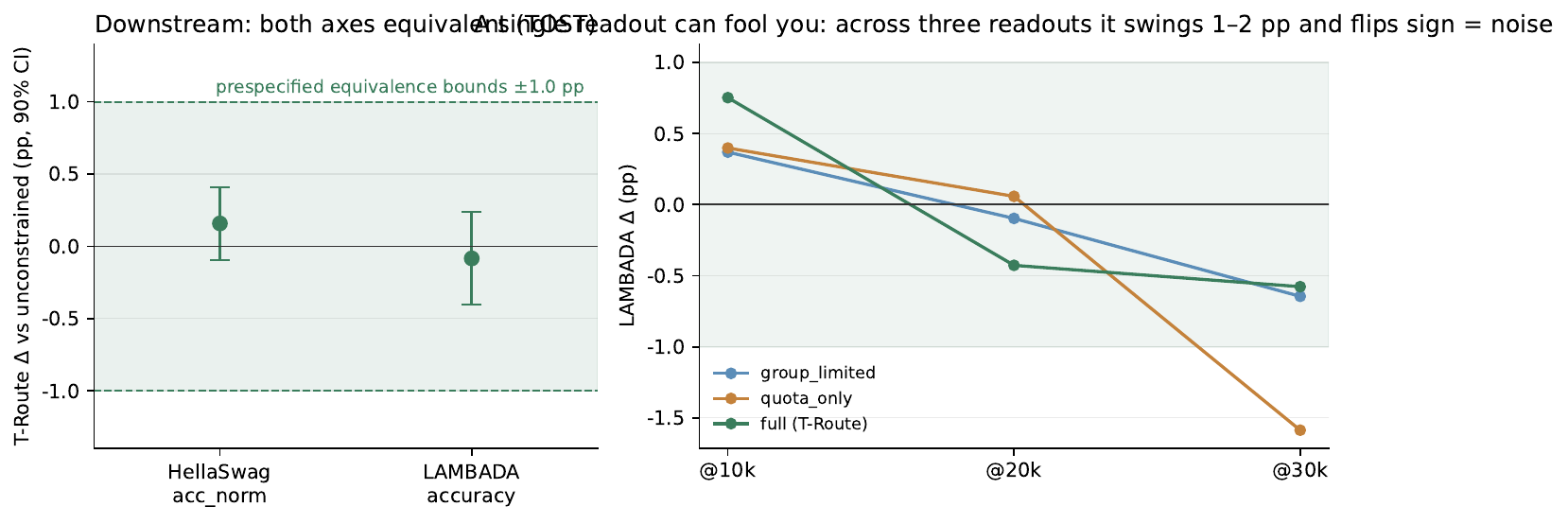}
\caption{Downstream axis. Left, the conjunction's delta against the unconstrained
control on the two protocol benchmarks, in percentage points with reported 90\%
confidence intervals, against the prespecified equivalence band of $\pm 1.0$ points. Right,
LAMBADA read at three checkpoints for each constrained mode.}
\label{fig:downstream}
\end{figure}

Downstream accuracy was scored under two one-sided tests at the prespecified bound of
$\pm 1.0$ percentage points~\cite{schuirmann1987tost,lakens2017equivalence}. The
recorded analysis reports equivalence on both benchmarks: HellaSwag at $+0.158$ with
interval $[-0.095, +0.411]$
over 3000 questions, and LAMBADA at $-0.084$ with $[-0.406, +0.238]$ over 5153
questions pooled across three checkpoints. Two qualifications belong with that verdict.
The raw per-seed/checkpoint results, pairing map, and variance/covariance estimator
behind those intervals are not all recorded in the artifact, so a reader cannot
independently reconstruct the tests. We therefore treat equivalence as a reported
result with incomplete reproducibility rather than as a fully auditable gate. The
pooling is also not a convenience. Read at 30k alone, LAMBADA puts the
conjunction at $-0.577$ with an interval touching the bound, which would have been
reported as a measurable loss; across three checkpoints each mode's estimate swings one
to two points and changes sign, a swing larger than any single estimate. At four seeds
this benchmark does not resolve effects below a point, so the axis carries an
equivalence claim and no point cost.

\begin{figure}[t]
\centering
\begin{subfigure}{0.635\textwidth}
\includegraphics[width=\textwidth]{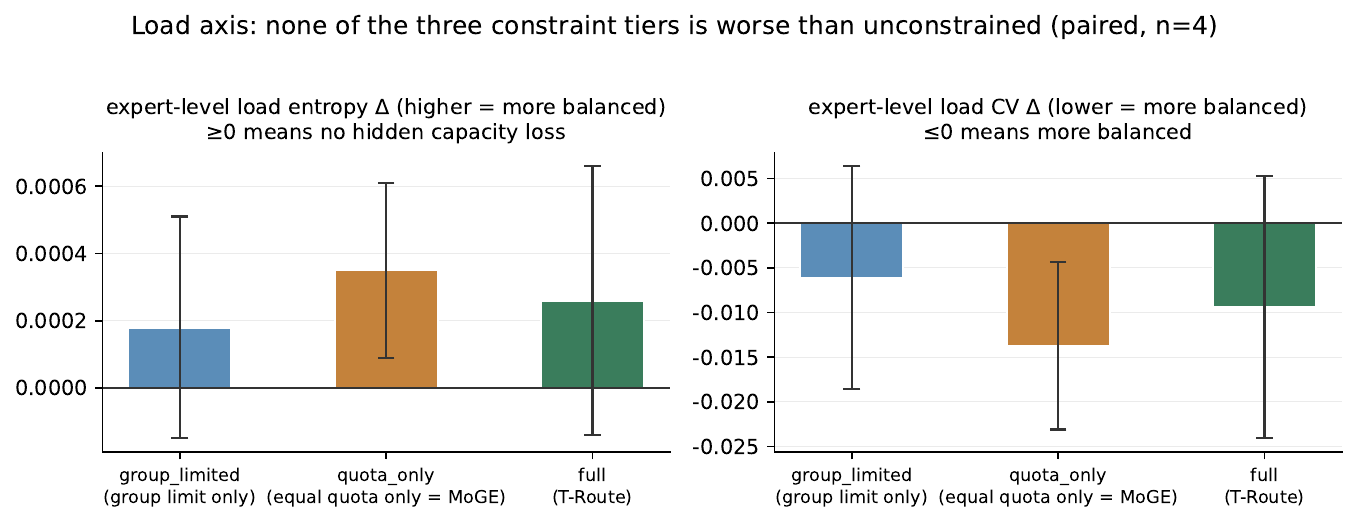}
\caption{Load, arm means of the paired deltas at $n = 4$, with across-seed spread.}
\end{subfigure}
\hfill
\begin{subfigure}{0.345\textwidth}
\includegraphics[width=\textwidth]{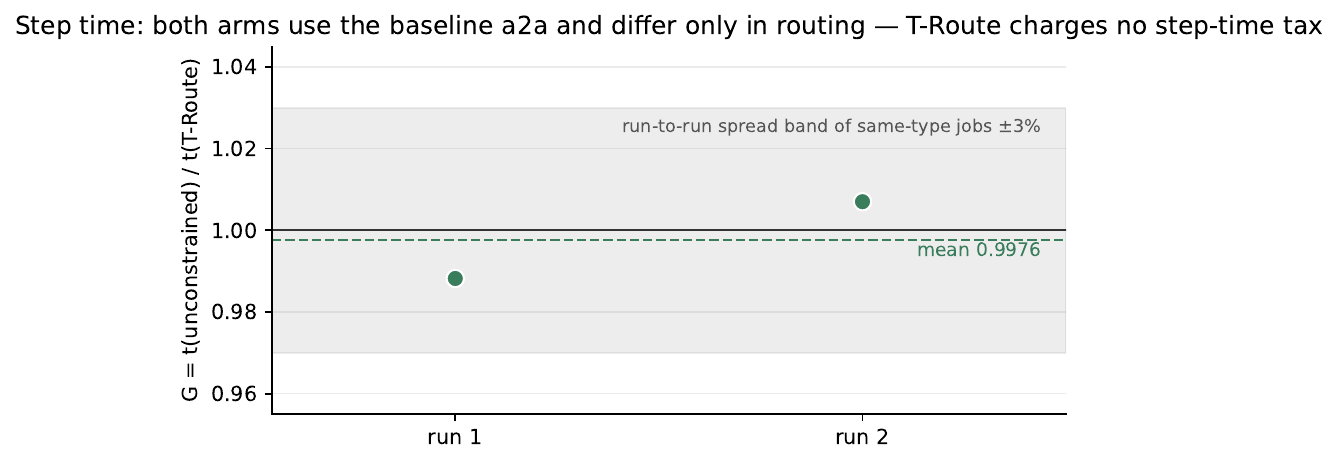}
\caption{Step time, two runs of one configuration.}
\end{subfigure}
\caption{Left, expert-level load entropy (higher is more balanced) and expert-level
load coefficient of variation (lower is more balanced), each as a difference against
the unconstrained control paired by seed. Right, the ratio of unconstrained to
constrained step time on two runs, against the assumed run-to-run band. The right
panel's axis label reuses the symbol $G$ for this routing-arm ratio, which is not the
one-hop to two-hop $G$ of Section~\ref{sec:problem}.}
\label{fig:loadstep}
\end{figure}

The load and step-time axes had no noninferiority margin, so their observed directions
establish less than the quality axis does. Expert-level load entropy is no lower
under any constrained mode, at $+0.00018$, $+0.00035$ and $+0.00026$ against the
control, and the coefficient of variation is no higher, at $-0.0061$, $-0.0137$ and
$-0.0094$; the spreads printed beside these in the artifact are not identified as any
particular dispersion and are read here only as scatter. The direction rules out a
hidden capacity loss traded for structure rather than establishing an improvement. Step
time was measured with both arms on the identical baseline one-hop all-to-all,
differing only in routing, over two runs of one configuration at 0.9882 and 1.0070,
mean 0.9976, inside the one to three percent run-to-run spread the artifact assumes for
jobs of this shape. Two runs of one configuration is the thinnest evidence in this
section.

Four boundaries hold on all of it. The geometry is one point, $q = 2$ at eight groups
of sixteen, and nothing here says what the constraints cost at a tighter group cap,
which is the direction Section~\ref{sec:applicability} shows the dispatch prefers. The
model is 13.14B parameters, some fifty times smaller than the models the dispatch
question is usually asked about. Four seeds resolve the quality axis and do not resolve
the downstream one. And the endpoint is validation loss on a holdout shard, which is
the standard proxy and not the thing anyone deploys. Subject to those four, an adopter
can evaluate the bounded per-token traffic envelope separately from the transport, but
must revalidate quality at its own geometry and scale.

\section{End-to-End Test of the Criterion on Seven Geometries}
\label{sec:e2e}

\begin{figure}[t]
\centering
\includegraphics[width=0.98\textwidth]{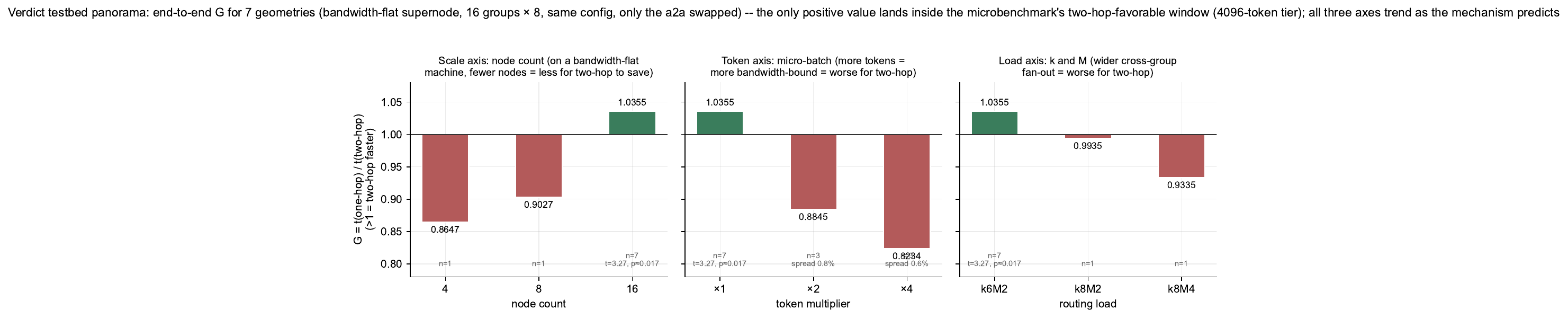}
\caption{Seven geometries, identical configuration, only the all-to-all
implementation swapped. $G$ is the ratio of one-hop to two-hop step time, so values
above 1 favour two-hop.}
\label{fig:verdictbed}
\end{figure}

Machine A sits at hierarchy ratio 1.03 (Table~\ref{tab:machines}), below every
effective breakeven of Section~\ref{sec:breakeven}, so the criterion predicts that
two-hop loses there on bytes. The fixed-cost condition of Section~\ref{sec:profile},
$\alpha(N_g) + \alpha(R) < \alpha(EP)$, does clear on this machine, at 0.157 plus
0.111 against 0.378~ms (Section~\ref{sec:model}), but it clears by about a tenth of a
millisecond per call pair, so two hops save almost nothing on fixed cost and nothing
on that side offsets the verdict on bytes. Six of the seven measured geometries have a
direction the criterion can be scored on. Five of those agree with that verdict
(Figure~\ref{fig:verdictbed}), and the sixth, the base tier, measures above 1 and is a
measured miss rather than a rounding. The seventh sits inside the run-to-run spread of
unity, and Section~\ref{sec:gatereg} excludes it from the sign count for that reason.

The two arms differ only in which all-to-all implementation the dispatch and combine
calls resolve to. Eight cards per group, hidden width 2048, sequence length 4096,
global batch 512, bf16 activations and 19 mixture-of-experts layers in a 20 layer
stack are held fixed across all seven, which are corpus G1 of
Section~\ref{sec:corpora}. The swapped path is issued often: at the base tier each step
makes 152 forward-shaped dispatch calls, half of them the recompute replay, and 76
backward ones, with as many again on combine. Each arm's step time is the median of ten
steady-state steps after the first 300 are discarded, which excludes warmup and limits
the influence of single-step outliers; across runs this bed reports a mean rather than
the pooled median of Section~\ref{sec:protocol}.

The seven geometries make nine bars in Figure~\ref{fig:verdictbed}, because the base
tier is the shared anchor of all three axes and is drawn on each. The scale axis varies
node count, the token axis the size of a micro-batch, and the load axis $k$ and then
$M$, which are the quantities Equation~\ref{eq:twohop} prices. No axis isolates a
single factor. Global batch is held fixed while expert parallelism shrinks with node
count, so the number of micro-batches per step, and with it the all-to-all calls per
step, rises steeply along the scale axis, and the first step away from the anchor on
the load axis changes $k$ before it changes $M$.

All three axes fall away from the base tier, and no model this paper adopts reproduces
that ordering: the Tier-2 row of Section~\ref{sec:validation} predicts a ratio above 1
at all seven geometries and places none of the six holdouts inside the tolerance band,
and the overlap families that do recover the directions were enumerated after that
tier failed and none was adopted. Increasing the size of a micro-batch makes two-hop
worse, from $G = 1.0355$ at the base size through $0.8845$ at twice it to $0.8234$ at
four times it; the driver is not established here, since on this bandwidth-flat
machine the modelled difference in bytes is a small remainder beside the arrival chain,
and Section~\ref{sec:validation} reads the exposure varying along this axis as a
property of the schedule. Reducing node count makes it worse, $0.9027$ at 8 nodes and
$0.8647$ at 4, along an axis on which Hop A spans fewer ranks while the deduplication
quota $q$, fixed by $k$ and $M$, does not change. Widening cross-group fan-out makes it
worse, $0.9935$ at $M=2$ against $0.9335$ at $M=4$, both at $k=8$, which halves $q$
from 4 to 2, and $\rbe$ rises as $q$ falls (Equation~\ref{eq:breakeven}). These are
ratios of whole training steps rather than of isolated calls: the one-hop arm takes
4420.4~ms per step at the base tier and 11712.0~ms at four nodes.

The seven points are not equally supported, and Figure~\ref{fig:verdictbed} prints
each run count. The base tier is the mean of seven runs; beyond it and the two
three-run geometries Section~\ref{sec:protocol} takes its spread from, the remaining
four, both load-axis points away from the anchor and both reduced node counts, were
run once. Three of those four lose by margins many times the spread this bed reports on
the token axis, so their direction is unlikely to be an artefact of a single reading,
though carrying that spread to 8 and 4 nodes is untested. The fourth is $k=8$ with
$M=2$, the undirected point.

The geometry that measures above 1 is the base tier, which is also the calibration
geometry and so the one point Tier-2 reproduces by construction rather than predicts.
It sits 3.55\% above 1, with $t = 3.27$ and $p \approx 0.017$, and the spread the rule
of Section~\ref{sec:falsification} compares against, 0.6\% to 0.8\% on this bed
(Section~\ref{sec:protocol}), does not cover that margin. By that rule the criterion is
refuted on this machine. Why this geometry goes positive is not established here. The
microbenchmark of corpus C1 still favours two-hop at this tier of 4096 tokens per rank,
at 1.366~ms one-hop against 1.217~ms two-hop, in a pure-communication form carrying
neither the splits exchange nor the arrival chain, and its crossover falls between that
tier and the next one measured, at the token count the doubled micro-batch size runs
at. C1 was collected at one routing configuration and one world size and stops at that
next tier, so it is silent about the other five points. A phase-level record that the
two-hop arm's combine runs faster also exists, but phase spans do not compose to step
time here, and the one constant fitted at this geometry is solved from the ratio itself
(Section~\ref{sec:validation}), so neither bears on the ratio's sign.

\section{Limits of the Model}
\label{sec:limits}

\subsection{Pricing the expert matmuls}
\label{sec:compute}

Whether a square roofline~\cite{williams2009roofline}, its ceiling measured on the
machine rather than taken from vendor peak~\cite{lo2014rooflinetoolkit}, can price an
expert matmul depends on how the curve is indexed by a non-square shape, and the three defensible rules, smallest dimension,
geometric mean and row count, diverge by up to $2.25\times$. That divergence is why an
earlier version of this work reported a bracket rather than a number, and why it read
the 43\% efficiency at square-1024 as evidence that narrow experts perform their work
less efficiently rather than merely performing less of it.

On a bf16 square GEMM roofline, achieved throughput peaks at 327~TFLOPS at size 4096
and falls to 141~TFLOPS at 1024, 43\% of the machine's own peak, with a non-monotone
dip through 8192 to 12288 that both campaigns reproduce (R1, 256 samples per size,
two campaigns). Eight expert-FFN shapes on each of two nodes, with same-session square
references that reproduce the shipped curve to 0.4\%, resolve the index question
directly (R2), and the three rules are not equally wrong
(Table~\ref{tab:indexrules}). The bracket was not a statement about what the data
could support but a statement about not having looked.

\begin{table}[t]
\centering
\small
\begin{tabular}{@{}lrrr@{}}
\toprule
Index rule & median $|$error$|$ & bias & worst \\
\midrule
smallest dimension & 19.0\% & $-19.0\%$ & 37.1\% \\
row count & 11.2\% & $-11.2\%$ & 37.1\% \\
\textbf{geometric mean} & \textbf{6.4\%} & $-6.4\%$ & \textbf{10.9\%} \\
\bottomrule
\end{tabular}
\caption{Three ways of indexing a square roofline by a non-square shape, scored against
the eight measured expert-FFN GEMMs of R2 on two nodes.}
\label{tab:indexrules}
\end{table}

A GEMM whose smallest dimension is 1536 measures 310 to 328~TFLOPS, not the
206 the square curve implies at that size. \textbf{The dip at square-1024 is a total-work
effect, not a shape effect}: a square 1024 matmul is 2.1~GFLOP and never fills the
machine, while an expert FFN of $1536 \times 2048 \times 5504$ is 34.6~GFLOP and
does. \emph{The earlier reading was wrong and we withdraw it}: narrow experts do not
perform their work less efficiently on account of being narrow. What survives is
narrower and still useful: efficiency falls when the work \emph{per call} is small,
which is a statement about expert count and micro-batch size rather than about expert
width on its own. The measurement that settles this is one an earlier version of this
paper listed among the things it had not done.

All three rules under-predict, because
a square curve is a systematically pessimistic prior for a large non-square GEMM, but
the geometric mean is within 6.4\% and the model uses it. We leave the residual bias
uncorrected: eight points on one machine is not enough to fit a correction onto, and
predicting compute slightly slow understates communication's share, which is the safe
direction for a quantity we only ever use as an upper bound.

\subsection{A machine the fast-domain abstraction does not describe}
\label{sec:nonuniform}

Equation~\ref{eq:call} gives a fast domain one bandwidth, $\bfast$, and the profile
of Section~\ref{sec:profile} asks whether the hierarchy ratio clears a threshold.
Both presume the domain is uniform: that any pair of accelerators inside it sees the
same link, so a collective inside the domain has one bandwidth to be priced at. On
the machines in Table~\ref{tab:predict} that presumption holds, because an
NVLink domain behind a switch and a torus slice are both regular by construction.

It does not hold everywhere, and Frontier is the instructive counterexample. Its
node carries eight accelerator dies, and the facility's own documentation states
that the interconnect topology is not fully connected and that peak bandwidth
between dies on different packages ranges from 50 to 100~GB/s depending on how many
links a given pair happens to have, while the two dies sharing a package see
200~GB/s. There is no single fast-side bandwidth on such a node. A figure obtained
by summing every link incident on a die describes a capacity no collective can
reach, since a ring across the eight dies is gated by its weakest edges, and the
resulting hierarchy ratio would be several times the one a collective actually
experiences. That is not a small error in a cell; it places the machine in the wrong
region of the criterion.

The right response is for the model to decline rather than to answer. A cost model
that reports a ratio for a non-uniform domain is not conservative, it is confidently
wrong, and this is the second place in this paper where the honest output is a
refusal rather than a number. We therefore state the precondition explicitly:
$\bfast$ is defined only where a domain is regular enough that one bandwidth
describes every pair inside it, and machines whose fast side is a heterogeneous
graph need a model with an edge-level cost term that this one does not have.

\subsection{What the model does not predict}

The communication share the model exposes is an \emph{upper bound} taken under the
assumption of no overlap, which is why Section~\ref{sec:model} adds compute to
communication nowhere in this paper: the two overlap by an amount our measurements do
not resolve, the same gap that keeps Tier-2 locked. The bound is useful for ranking
geometries and for identifying regimes where communication cannot dominate, and it is
labeled a bound rather than a prediction. Three further capabilities are absent
because a gate outcome removed them rather than because we were cautious: step-level
prediction (Section~\ref{sec:tier2}), any statement above world 128
(Section~\ref{sec:coverage}), and any reading of the arrival chain model as more than
a lower bound below 8192 rows or above $H = 2048$
(Section~\ref{sec:chainscaling}).

\section{Related Work}
\label{sec:related}

\paragraph{Expert-parallel MoE systems.} The sparsely gated MoE
layer~\cite{shazeer2017moe} and its expert-parallel dispatch and combine
formulation~\cite{lepikhin2021gshard,fedus2022switch} put an all-to-all in the inner
loop of training. Systems work since has attacked that all-to-all from several
directions: adaptive parallelism and pipelining~\cite{hwang2023tutel}, contention
with the data-parallel allreduce~\cite{li2023lina}, moving experts instead of
tokens~\cite{liu2023janus}, stacks combining several of
these~\cite{shi2024schemoe,jin2026megascalemoe}, one of them a production system, and
inference-side hierarchical decomposition of the same
collective~\cite{rajbhandari2022deepspeedmoe}. The tensor and pipeline parallelism
such systems compose with is standard~\cite{shoeybi2019megatron}. A
parallel line attacks the compute side instead, dropping the capacity constraint in
favour of block sparse GEMM~\cite{gale2023megablocks}. Our
target is narrower and complementary: not a faster all-to-all but a way to decide,
before building one, whether a particular restructuring pays on a particular machine.

\paragraph{Hierarchical dispatch and node-limited routing.} The idea of sending one
copy across a slow boundary and fanning out behind it is old in HPC. Hierarchical
all-to-all as node-local gather, inter-node exchange, node-local
scatter~\cite{traff2014hierarchical} and node-aware allreduce that separates the
intra- from the inter-node phase and removes duplicate messages across the node
boundary~\cite{bienz2019nodeaware} are the direct ancestors. In MoE it appears as
device-limited routing~\cite{deepseekv2} and then as node-limited routing with
explicit forwarding from the slow fabric onto the fast
one~\cite{deepseekv3,deepseekv3hw}, as grouped experts constrained to balance across
device groups~\cite{tang2025pangumoge}, as redundancy-bypassing
dispatch~\cite{xmoe2025}, and in a device-initiated communication
library~\cite{deepep2025}. Closest to this paper is HierMoE~\cite{lin2025hiermoe},
which builds theoretical models of token deduplication under different topologies
and reports speedups. Section~\ref{sec:design} states what we do with such a model
instead of building a faster one.

\paragraph{Communication cost models.} Equation~\ref{eq:call} is the Hockney
form with its half-performance message size~\cite{hockney1994}, a specialisation of
the half-performance parameter family the same author introduced for supercomputer
benchmarking~\cite{hockney1991}. LogP~\cite{culler1993logp} splits that startup term into
overhead, latency and a per-message gap, which matter for trains of small messages;
LogGP's per-byte gap~\cite{alexandrov1995loggp} is the long-message slope our
bandwidth term already carries. We do not model the overhead and latency split,
because the MoE regime is bandwidth-dominated at the sizes that decide the outcome,
and we say so as a modelling choice rather than an oversight.
Cost expressions per collective and their use in algorithm
selection~\cite{thakur2005mpich,ricogallego2019survey} are the direct precedent for
predicting a collective's time from a small number of fitted constants. FasterMoE models MoE
all-to-all cost directly, from measured link bandwidths~\cite{he2022fastermoe}. Our methodological debt on how to report such
measurements honestly is to Hoefler and Belli~\cite{hoefler2015benchmarking}, whose
prescriptions on repetition, confidence intervals, and distinguishing real from
chance improvements are what our gate thresholds and pooled-median targets implement.

\paragraph{Compute-side modelling.} Section~\ref{sec:compute} borrows the
empirical-ceiling discipline of the roofline model~\cite{williams2009roofline}, with
the ceiling measured rather than taken from vendor peak~\cite{lo2014rooflinetoolkit};
it uses the compute roof alone and no operational-intensity axis, so it is a
throughput curve indexed by problem size rather than a roofline in the original
sense. Our
earlier refusal to convert a square-GEMM roofline into an expert-matmul time rested
on the same premise as the tall-and-skinny literature, that such shapes deliver a
fraction of what a square roofline predicts, which has performance models of its
own~\cite{ernst2021tallskinny,rivera2021tsm2x}. Our measurements do not reproduce
that penalty at expert-FFN sizes: the shapes we ran are tall and skinny by aspect
ratio but large in total work, and they reach the square peak. The distinction that
matters on this hardware is total work per call rather than aspect ratio. Separately,
transformer step time is data-movement bound in ways operator-level FLOP counting
misses~\cite{ivanov2021datamovement}, which is an independent reason not to price a
step from GEMM shapes alone.

\paragraph{Simulators for distributed training, and how they validate.} Analytical
co-design tools~\cite{isaev2023calculon}, hierarchical network
simulators~\cite{won2023astrasim2}, replay-based
diagnosis~\cite{hu2022dpro}, and full training or inference
simulators~\cite{duan2024proteus,wang2025simai,agrawal2024vidur} all report accuracy
against measurement, typically a single aggregate error figure in the single-digit
percent range. Trace-driven analytic simulation validated against real runs goes back
at least to LogGOPSim~\cite{hoefler2010loggopsim}. What we add is not a lower error
number but a different reporting contract: multiple gates with thresholds fixed in
advance, an independent benchmark family as a second gate, a gate that fails and
blocks the extrapolation it would license, and, unlike a single aggregate error
figure, a contract under which a gate outcome constrains what the model may be used
for. Our step-level gate fails, and the consequence is that this paper makes no
step-level prediction at all.

\section{Conclusion}

On a target machine, hierarchical MoE dispatch can be screened at the
communication-call level if the target's constants are measured at the workload's
message sizes and the communication-level gates pass. The present data do not license
a step-time or training-throughput verdict: the step-level gate fails.

The corrected Hop-A accounting gives effective breakevens of 3.98 for the measured
PyTorch chain, 1.49 for a hypothetical fused target, and 1.10 at zero implementation
overhead at the reference geometry. The arrival chain is therefore the dominant local
sensitivity in this calibration, not a demonstrated cross-platform cause. T-Route
provides a bounded per-token fan-out and quota; its validation-loss effect is small
but nonzero, and its downstream equivalence record is not independently
reconstructible from the released artifact. Machine B shows that the same functional
form can retain acceptable fit after machine-specific refitting on C3; it is not an
out-of-sample transfer test.

Coverage is the main limitation. Only machine A has a measured hierarchy ratio, 1.03,
and machine B's ratio is unresolved. Every row above 1.03 in the applicability table
is therefore a synthetic ratio sensitivity, not an observation of a hierarchical
cluster. A machine measured in that regime, plus a communication-level holdout and
an overlap-aware step instrument, are the measurements needed to strengthen the claim.

\section*{Reproducibility}
\label{sec:repro}

The model is parameterised by five quantities that must be measured on a target:
$\alpha$ per world size,
$\bfast$, $\bslow$, $\xhalf$, and $c_{\mathrm{chain}}$ for the arrival chain
implementation in use. Only $\xhalf$ needs a wide size sweep, and
Table~\ref{tab:constants} shows it survives being borrowed across machines while
passing Tier-1; the remaining four each come from a single microbenchmark.
Re-calibration on a new machine therefore costs far less than the build the criterion
is deciding about, which is why the constants in Table~\ref{tab:constants} are
reported as properties of two specific machines rather than as defaults.

We publish the distilled constants and gate code but not the raw measurement sweeps or
the complete downstream estimator inputs. That is enough to run the procedures against
new measurements, but not to independently re-derive every reported gate outcome from
our data. The artifact is therefore executable but not a full measurement
reproduction package.

Every figure in this paper is produced by the same script that produces the
repository's copy of it, so a number in the paper cannot drift from the same number in
the code, and rerunning a script reproduces the file byte for byte. Model figures are
computed live by calling the modules the text cites; measurement figures carry their
numbers inline, and the script is then the record of what was measured.

The cost model, the calibration constants, the validation gates, the figure
generation, the applicability profile, and the checks that enforce the capabilities
removed in Section~\ref{sec:gatereg} are at
\url{https://github.com/weich97/TerraceMoE-simulator} under Apache-2.0. The gates
there include the two that fail, and the criterion is executable rather than described
only in prose, so a replication scores its own targets against the same thresholds
instead of against our reported outcomes.

\clearpage
\bibliographystyle{plain}
\bibliography{refs}

\end{document}